\documentclass[12pt,a4paper,JHEP3]{article}

\pdfoutput=1
\usepackage{hhline}
\usepackage[table,dvipsnames]{xcolor}
\usepackage{jheppub}
\usepackage{slashed,cancel,lscape,caption,array,graphicx,subcaption}
\usepackage[utf8]{inputenc}
\usepackage{amsmath}
\usepackage{nicefrac}  
\usepackage{mathtools}
\usepackage{multirow}
\usepackage{booktabs} 
\usepackage{chngcntr}
\usepackage{nicefrac}
\usepackage{cleveref}
\usepackage[normalem]{ulem}
\usepackage{setspace} 
\usepackage{makecell} 

\newcommand{\wilson}{C}
\newcommand{\wilsonZ}{C^{\rm SSC}}
\newcommand{\Z}{\mathcal{Z}}

\newcommand{\probeScale}{\lambda}

\usepackage[compat=1.1.0]{tikz-feynman} 

\usepackage{tikz}
\usepackage{adjustbox}
\usetikzlibrary{decorations.pathmorphing}
\usetikzlibrary{decorations.markings}
\usetikzlibrary{positioning, shapes, snakes, arrows}
\usetikzlibrary{patterns}

\tikzset{
    wl/.style={line width=1pt},
    graviton/.style={line width=.8pt, -latex,decorate, decoration={snake, segment length=4pt,amplitude=1.8pt, pre length=.15cm, post length=.25cm}},
    worldlineStatic/.style={dotted, line width=1pt},
	worldline/.style={gray, line width=1pt},
	worldlineBold/.style={black, line width=.6pt},
	zUndirected/.style={line width=1pt},
	zParticle/.style={line width=1pt,postaction={decorate},decoration={markings,mark=at position .6 with {\arrow[#1]{latex}}}},
	zParticleF/.style={line width=1pt,postaction={decorate}},
	cscalar/.style={line width=1pt,postaction={decorate},decoration={markings,mark=at position .6 with {\arrow[#1]{latex}}}},
	cscalar2/.style={line width=1pt,postaction={decorate},decoration={markings,mark=at position .8 with {\arrow[#1]{latex}}}},
	photon/.style={line width =.8pt, decorate, decoration={snake, segment length=4pt, amplitude=1.8pt,  pre length=.1cm, post length=.1cm}},
	photonRed/.style={red, line width =.8pt, decorate, decoration={snake, segment length=4pt, amplitude=1.8pt,  pre length=.1cm, post length=.1cm}},
	cross/.style={cross out, line width =.8pt, draw=black, minimum size=2*(#1-\pgflinewidth), inner sep=0pt, outer sep=0pt},
cross/.default={4pt}
}

\newcommand{\jitze}[1]{\textbf{\textcolor{orange}{[Jitze: #1]}}}

\newcommand{\zenodo}{\href{https://doi.org/10.5281/zenodo.15681340}{\tt zenodo.org} \cite{zenodo} }

\DeclareFontFamily{OT1}{pzc}{} 
\DeclareFontShape{OT1}{pzc}{m}{it}{<-> s * [1.350] pzcmi7t}{}
\DeclareMathAlphabet{\mathpzc}{OT1}{pzc}{m}{it}

\def\cL{\mathcal{L}}

\def\cO{\mathcal{O}}

\def\cV{\mathcal{V}}

\def\eps{\epsilon}

\def\d{\mathrm{d}}
\def\D{\mathrm{D}}

\def\pat{\partial}
\def\mn{{\mu\nu}}
\def\ab{{\alpha\beta}}

\def\i\math

\def\bH{\hat{b}}

\def\d{\mathrm{d}}
\def\eps{\epsilon}

\renewcommand{\i}{\ensuremath{\mathrm{i}}}
\renewcommand{\d}{\ensuremath{\mathrm{d}}}

\def\nn{\nonumber}

\def\eqn#1{eq.~\eqref{#1}}

\newcommand{\vev}[1]{\langle #1\rangle}

\newcommand{\be}{\begin{equation}}
\newcommand{\ee}{\end{equation}}
\newcommand{\ba}{\begin{align}}
\newcommand{\ea}{\end{align}}
\ifx\genfrac\sdflkaj\else\fi
\newcommand{\sfrac}[2]{{\textstyle\frac{#1}{#2}}}

\newcommand{\xd}{\dot x}

\def\centerarc[#1](#2)(#3:#4:#5){ \draw[#1] ($(#2)+({#5*cos(#3)},{#5*sin(#3)})$) arc (#3:#4:#5); }

\begin{document}

\begin{flushright}
\begingroup\footnotesize\ttfamily
	HU-EP-25/22-RTG
\endgroup
\end{flushright}

\vspace{15mm}

\begin{center}
{\LARGE\bfseries 
Spinning the Probe in Kerr with WQFT
\par}

\vspace{15mm}

\begingroup\scshape\large 
     Jitze Hoogeveen,${}^{1}$ 
	Gustav Uhre Jakobsen,${}^{1,2}$  
	and Jan~Plefka${}^{1}$
\endgroup
\vspace{5mm}
					
\textit{${}^{1}$Institut f\"ur Physik, Humboldt-Universit\"at zu Berlin, 10099 Berlin, Germany} \\[0.25cm]
\textit{${}^{2}$Max-Planck-Institut f\"ur Gravitationsphysik
(Albert-Einstein-Institut), 14476 Potsdam, Germany } \\[0.25cm]

\bigskip
  
\texttt{\small\{jitze.hoogeveen@hu-berlin.de, gustav.uhre.jakobsen@physik.hu-berlin.de, 
jan.plefka@hu-berlin.de\}}

\vspace{10mm}

\textbf{Abstract}\vspace{5mm}\par
\begin{minipage}{14.7cm}
We investigate the gravitational scattering of a spinning probe mass in a Kerr background using the worldline quantum field theory approach. This corresponds to the leading term (0SF) in the gravitational self-force expansion for the spinning two-body problem with large mass hierarchy. By reformulating the geodesic and Mathisson--Papapetrou--Dixon equations as a recursive
Berends-Giele type equation known from multi-gluon scattering, we develop a novel integration-by-parts formalism on the worldline that enables systematic computation of scattering observables—specifically the impulse and spin kick—to arbitrary orders in Newton's constant and spin. Here the transition to a position space
formalism is key. We present explicit results up to and including the physical 7PM order, thereby incorporating all relevant higher-spin and higher-curvature terms on the worldline, advancing beyond previous calculations. This work represents an initial step  to reconceptualise the gravitational self-force expansion through
worldline quantum field theory.
\end{minipage}\par

\end{center}
\setcounter{page}{0}
\thispagestyle{empty}
\newpage
 
\tableofcontents
\clearpage
\section{Introduction}
The gravitational two-body problem --- the classical dynamics of two massive point particles interacting through their gravitational attraction --- is fundamental, conceptually simple and
 yet surprisingly complex. While the Newtonian case is integrable, the fully relativistic Einsteinian problem presents significant challenges. Depending on initial conditions, bound motion results in inspiraling mergers (now routinely observed for black holes and neutron stars in gravitational wave detectors \cite{KAGRA:2021vkt}), while unbound scenarios produce scattering events emitting gravitational ``Bremsstrahlung'' bursts that are 
 potentially observable by future detectors like LISA \cite{LISA:2017pwj,Amaro-Seoane:2012lgq,Hopman:2005vr,McCart:2021upc}

Due to the complexity of Einstein's equations, exact solutions remain unattainable. Consequently, one either resorts to numerical approaches \cite{Pretorius:2005gq,Damour:2014afa,Boyle:2019kee} or considers approximate analytical schemes with identified small parameters. Three perturbative frameworks provide analytical control: (i) the post-Newtonian expansion (PN) particularly suited for bound motion scenarios, assuming non-relativistic velocities and weak fields ($v^2\sim Gm/r\ll 1$) \cite{Blanchet:2013haa,Porto:2016pyg,Levi:2018nxp}, 
(ii) the post-Minkowskian expansion (PM) intensively used for the scattering scenario, assuming only weak fields 
($Gm/b\ll 1$)~\cite{Westpfahl:1985tsl,Damour:2017zjx,Bern:2019nnu,Damour:2019lcq,Kalin:2020fhe,DiVecchia:2021bdo,Kalin:2020mvi,Mogull:2020sak,Bern:2021yeh,Dlapa:2021vgp,Dlapa:2022lmu,Damgaard:2023ttc,Jakobsen:2023hig,Kosower:2022yvp,Bjerrum-Bohr:2022blt,Buonanno:2022pgc,DiVecchia:2023frv}, 
and (iii) the gravitational self-force expansion (SF) for two body encounters with large mass hierarchies, expanding in the mass ratio $m/M$ of the light (or probe) body $m$ and the heavy body $M$ \cite{Mino:1996nk,Poisson:2011nh,Barack:2018yvs,Gralla:2021qaf}. The latter is particularly relevant for extreme mass ratio inspirals that will be observable with LISA. Spin significantly influences all observables in binary processes, making it very relevant and contributes to PM/PN expansion orders\footnote{As the spin $S$ of a black hole is bounded by $Gm^2$, powers in spin effectively increase the order in $G$. 
Therefore a contribution of order $G^{n} S^{m}$ is of the ``physical'' $(n+m)$th PM order.}.

The leading term in the self-force expansion is the 0SF or probe limit. This is equivalent to solving the geodesic and Mathisson--Papapetrou--Dixon equations (for the spinning case) for a (spinning) probe mass $m$ moving in a fixed Schwarzschild (or Kerr) background of mass 
$M$ ignoring back reaction\footnote{Note that in the self-force literature higher orders in spin of the probe are attributed to
higher SF orders. We do not adopt this counting here and attribute the SF order solely to
the powers in the symmetric mass ratio $\nu=mM/(m+M)$ (ignoring the mass scaling of the black holes spins --- more detail in the main text).}. 
Using traditional general relativity (GR) methods exact solutions for spinless probes in Kerr backgrounds have been
established \cite{Carter:1968ks,Dixon:1970zza,Sharp:1979sqa,Mino:2003yg,Hackmann:2013pva,Damgaard:2022jem}, and extended to lower orders in probe spin in combination with the PM expansion \cite{Rudiger:1981,Skoupy:2024uan,Gibbons:1993ap,Compere:2023alp,Hoogeveen:2023bqa}. 

Recent years have witnessed remarkable progress in the PM expansion for the scattering problem
through synergistic applications of perturbative quantum field theory methods to this classical gravitational 
problem~\cite{Goldberger:2004jt,Bjerrum-Bohr:2022blt,Kosower:2022yvp,Buonanno:2022pgc}. These employ scattering amplitudes ~\cite{Neill:2013wsa,Luna:2017dtq,Kosower:2018adc,Cristofoli:2021vyo,Bjerrum-Bohr:2013bxa,Bjerrum-Bohr:2018xdl,Bern:2019nnu,Bern:2019crd,Bjerrum-Bohr:2021wwt,Cheung:2020gyp,Bjerrum-Bohr:2021din,DiVecchia:2020ymx,DiVecchia:2021bdo,DiVecchia:2021ndb,DiVecchia:2022piu,Heissenberg:2022tsn,Damour:2020tta,Herrmann:2021tct,Damgaard:2019lfh,Damgaard:2019lfh,Damgaard:2021ipf,Damgaard:2023vnx,Aoude:2020onz,AccettulliHuber:2020dal,Brandhuber:2021eyq,Bern:2021dqo,Bern:2021yeh,Bern:2022kto,Bern:2023ity,Damgaard:2023ttc,Brandhuber:2023hhy,Brandhuber:2023hhy,Brandhuber:2023hhl,DeAngelis:2023lvf,Herderschee:2023fxh,Caron-Huot:2023vxl,FebresCordero:2022jts,Bohnenblust:2023qmy}
 or worldline approaches~\cite{Kalin:2020mvi,Kalin:2020fhe,Dlapa:2021npj,Kalin:2022hph,Dlapa:2022lmu,Dlapa:2023hsl,Mogull:2020sak,Jakobsen:2021zvh,Jakobsen:2022fcj,Jakobsen:2022zsx,Jakobsen:2022psy,Jakobsen:2023ndj,Jakobsen:2023hig,Haddad:2024ebn,Liu:2021zxr,Mougiakakos:2022sic,Riva:2022fru,Shi:2021qsb,Bastianelli:2021nbs,Comberiati:2022cpm,Comberiati:2024uuc,Wang:2022ntx,Ben-Shahar:2023djm,Bonocore:2021qxh,Bonocore:2024uxk,Bonocore:2025stf}. For scattering observables such as the scattering angle and radiated energy, the present state-of-the-art is at the physical 5PM order: the four-loop ($G^5$) spin-less case \cite{Driesse:2024xad,Driesse:2024feo} (up to 1SF order), the three-loop up to linear order in spin  ($G^4S$) \cite{Jakobsen:2023ndj,Jakobsen:2023hig}, the two-loop  up to quartic order in spin ($G^3 S^{4}$)\cite{Jakobsen:2022fcj,Jakobsen:2022zsx,Akpinar:2025bkt} along with higher-order spin results at one-loop ($G^2S^{5}$) order \cite{Bern:2022kto}, and even all-order spin results at tree-level ($G S^{\infty}$)  \cite{Vines:2017hyw} and one-loop ($G^2S^\infty$)
\cite{Aoude:2023vdk}. 

In this work, we focus on spinning black hole (or neutron star) scattering in the probe or 0SF limit using the 
spinning worldline quantum field theory approach \cite{Mogull:2020sak,Jakobsen:2021zvh,Jakobsen:2022psy,Haddad:2024ebn}. We reformulate the geodesic and Mathisson--Papapetrou--Dixon equations into a Berends-Giele \cite{Berends:1987me} type recursive equation in worldline quantum field theory. An iterative solution technique 
is developed, employing a novel integration-by-parts (IBP) formalism on the worldline. 
Here, in the Kerr background scenario, the transition to a configuration space treatment of the problem proves superior to the standard momentum space approach employed for flat space
backgrounds. It enables a recursive computation of the impulse (change of momentum) and spin kick (change of spin) to, in principle, arbitrary orders in Newton's constant and the spin of both probe and background masses. We provide explicit results up to and 
including the physical 7PM order along with the code to run to higher orders in an accompanying \zenodo submission. Importantly, we include all relevant higher-spin (up to quartic order) and higher-curvature
(up to quadratic order) non-minimal extensions of the effective worldline action accounting for a generic
spinning (neutron star) probe.

Our work represents a first step toward reconceptualising the self-force expansion using quantum field theory techniques. Recent efforts have begun exploring this direction \cite{Kosmopoulos:2023bwc,Cheung:2023lnj,Cheung:2024byb,Akpinar:2025huz}. We envision significant benefits in combining exact general relativity knowledge with perturbative quantum field theory approaches to further advance understanding of the gravitational two-body problem.

\section{Spinning Worldline Quantum Field Theory}

\subsection{General Formalism and Probe Limit}\label{sec:GFPL}

Worldline quantum field theory (WQFT) \cite{Mogull:2020sak,Jakobsen:2021zvh,Jakobsen:2022psy,Haddad:2024ebn} has proven to be an efficient framework for 
quantifying the classical gravitational two-body problem -- in particular
 in the post-Minkowskian approximation. Here, following a worldline effective field theory
 approach \cite{Goldberger:2004jt,Kalin:2020mvi},
the black holes (BH) or neutron stars (NS) are modelled as point particles of mass $m$ -- this being
a valid approximation as long as their separation $r$ is large compared to their intrinsic
sizes, i.e.~$ r \gg Gm$. The trajectory  of a compact body is denoted by 
$x^{\mu}(\tau)$ with the worldline parameter $\tau$, while its spin is conveniently modelled through a set of complex vectors $\alpha^{\mu}(\tau)$ on the worldline \cite{Haddad:2024ebn}
that induce the object's spin tensor
\be\label{eq:STBO}
S^{\mu\nu}(\tau)=-2m\i\, \bar\alpha^{[\mu}(\tau)\alpha^{\nu]}(\tau)\, \, .
\ee
In this formalism the action for a spinning body of mass $m$ 
in a generalized proper-time gauge that moves
in a curved space-time background $g_{\mu\nu}$ takes the form \cite{Haddad:2024ebn}
%
\begin{align}\label{eq:sWQFTH}
S_{\rm WQFT}:=&
 -\int\!\d\tau\, \Bigl [ 
\sfrac{m}{2} g_{\mu\nu}\dot x^{\mu}\dot x^{\nu} + \i m \bar \alpha_{\mu} \frac{\D\alpha^{\mu}}{\d\tau}
\Bigr ] + S_{\rm nm}\, .
\end{align}
Here $\frac{\D\alpha^{\mu}}{\d\tau}= \dot\alpha^{\mu} + \Gamma^{\mu}{}_{\nu\rho}\xd^{\nu}\alpha^{\rho}$ is the covariant worldline derivative employing the Levi-Civita 
connection.

The coupling of spin to the space-time curvature is described by non-minimal terms in $S_{\rm nm}$ starting at quadratic order in spins --- the linear-in-spin term is universal, i.e.~identical for black holes and neutron stars.
The non-minimal terms capture a systematic expansion of higher-dimensional operators of increasing orders in spin and curvature in the sense of effective field theory~\cite{Haddad:2024ebn} which we will discuss in more detail in section~\ref{sec:nonminimalterms} below.

 The gravitational two-body problem may now be studied upon taking two copies of the action \eqn{eq:sWQFTH} for the masses $m$ and $M$  and trajectories 
 $x^{\mu}(\tau)$ and $X^{\mu}(\tau)$
 subject to the gravitational interaction
implemented by adding the Einstein-Hilbert and a gravitational 
gauge fixing term to the action.
One then performs a background field 
expansion of 
$\{x^{\mu}(\tau),X^\mu(\tau)\}$
and $g_{\mu\nu}$ about the non-interacting ($G=0$) situation:
straight line trajectories 
$\{b^\mu+v^\mu\tau,B^\mu+V^\mu\tau\}$
, constant spin $\{\alpha^{\mu}_{m},\alpha_{M}^\mu\}$, and flat Minkowski space-time $g_{\mu\nu}=\eta_{\mu\nu}$, introducing the deflection fields $\{z^{\mu}_{i}(\tau),\alpha_{i}^{\prime\mu}(\tau) \}$ (with $i=m,M$) on the worldline, and the graviton $h_{\mu\nu}(x)$ in the bulk. Next, one perturbatively 
integrates out these deflection modes in the path integral. 
The key feature of the WQFT
formalism is that the solution to the \emph{classical} equations of motion are
simply given by the tree-level one-point functions
$\langle x_{i}^{\mu}(\tau)\rangle$ and $\langle h_{\mu\nu}(x)\rangle$ of the WQFT. 
These are naturally computed in an expansion in powers of $G$, the weak gravity approximation or post-Minkowskian expansion.

A prime observable is the impulse, or change of momentum. 
It factorises into gauge invariant contributions according to their scaling in masses,
known as the gravitational self-force (SF) expansion. For the BH or NS of mass $m$ one has
\be
\Delta{p}^{\mu} = p^{\mu}(\tau)\Bigr |^{\tau= +\infty}_{\tau=-\infty}
\sim
m \sum_{n=1}^{\infty} 
(GM)^n
\sum_{k=0}^{n-1}
\left (\frac{m}{M}\right )^{k}\,  \Delta{p}_{(n,k)}^{\mu} \, .
\ee
The index $k$ refers to the self-force (SF) order. Upon permuting the particle labels
it is clear that one may relate the $\Delta p_{(n,k)}^{\mu}$ and $\Delta p_{(n,n-1-k)}^{\mu}$ contributions.

In the present work we focus on the
0SF or probe limit contributions $\Delta p^{\mu}_{(n,0)}$ which we shall establish
to arbitrary high, but fixed, orders in $G$ and spins. Physically these are dominant
for large mass ratios, i.e.~$m\ll M$. They amount to solving the equations
of motions for $x^{\mu}(\tau)$ emerging from \eqn{eq:sWQFTH} in a \emph{fixed} space-time
metric background, e.g.~the Kerr metric $g^{\rm Kerr}_{\mu\nu}$. 

Using the WQFT methodology this may be achieved upon performing a background
field expansion for the probe mass trajectory $x^{\mu}(\tau)$
\begin{equation}
 x^\mu(\tau)=b^\mu+v^\mu \tau+z^\mu(\tau)\, ,
\end{equation}
and the spin oscillators
\begin{equation}
  \alpha^\mu(\tau)=\alpha^\mu_{m}+\alpha'^\mu(\tau), 
  \quad\quad
   \bar \alpha^\mu(\tau)=\bar \alpha^\mu_{m}+\bar \alpha'^\mu(\tau)\, ,
\end{equation}
where $\alpha_m^\mu=\alpha^\mu(-\infty)$ is the initial state of $\alpha^\mu(\tau)$.
The dynamical worldline field content thus reads
\begin{equation}\label{eq:Zdef}
  \mathcal Z^\mu_{I}(\tau) =\{z^\mu(\tau),\alpha'^\mu(\tau),\bar \alpha'^\mu(\tau)\}, 
\end{equation}
upon which we integrate in the path integral to determine the tree-level one-point function $\vev{z^{\mu}(\tau)}$. Importantly, the background metric 
$g_{\mu\nu}=g^{\rm Kerr}_{\mu\nu}$ remains non-dynamical here, sourced just by the background black hole. We split it up into $G$-dependent and -independent parts as 
\begin{equation}\label{eq:metric_split}
  g^\mathrm{Kerr}_{\mu\nu}=\eta_{\mu\nu}+h^{\rm Kerr}_{\mu\nu}\, , 
  \qquad \eta_{\mu\nu}=\text{diag}(1,-1,-1,-1)\, ,
\end{equation}
working in the mostly minus metric convention.
 Crucially, $h_{\mu\nu}$ can be extracted to all orders in $G$ directly from the Kerr metric in position space. Probe dynamics thus naturally lends itself to a position space treatment. This contrasts the usual QFT preference for momentum space in handling dynamical two-body dynamics.
In this section, and in the following, indices are raised and lowered with the flat space-time metric $\eta_\mn$ unless otherwise stated.

Expanding around an asymptotically straight line trajectory, the 0SF impulse then follows from the quantity
\be
\Delta p^{\mu}_{\rm 0 SF}=  m\sum_{n=1}^{\infty} G^{n} M^{n}\,  \Delta{p}_{(n,0)}^{\mu}
= m\int_{-\infty}^{\infty}d\tau \vev{\ddot  z^\mu(\tau)} \, ,
\ee
with the probe mass $m$ and the Kerr-BH mass $M$.
The deflection may be expressed as
\begin{equation}\label{eq:pathintegral_definition}
   \langle z^\mu(\tau) \rangle=Z_{\rm WQFT}^{-1}\int \mathcal D[\mathcal Z_{I}^{\mu}] z^{\mu}(\tau)e^{\frac{i}{\hbar} S_{\rm WQFT}}
\end{equation}
which is to be computed at tree-level using  the Kerr metric in the worldline action $S_{\rm WQFT}$ and where
$Z_{\rm WQFT} = \int \mathcal D[\mathcal Z^{\mu}_{I}] e^{\frac{i}{\hbar} S_{\rm WQFT}}$ is the partition function.
An analogous expression
is obtained for the spin kick, i.e.~the change of spin $\Delta S^{\mu\nu}$, derivable from $\Delta\alpha=\vev{\alpha^{\prime \mu}(\tau=\infty)}$.

The background parameters of our expansion are identified as initial state parameters due to our use of causal (retarded) propagators.
The initial state spin tensor of the probe is
\begin{align}
  S^{\mu\nu}_{m}
  &=
  -2m\i\, 
  \bar\alpha_{m}^{[\mu}\alpha_{m}^{\nu]}
  \\
  &=
  m
  \eps^{\mn\ab}
  v_\alpha a_\beta
  \ .
\end{align}
Here we also introduced the initial state Pauli-Lubanski vector $a^\mu$ (with $\eps_{0123}=1$).
The Pauli Lubanski vector has dimensions of length and it is convenient to define a dimensionless counterpart $\chi^\mu$ through
\begin{align}\label{eq:chidef}
  a^\mu = \lambda \chi^\mu 
  \ ,
\end{align}
where we introduced a length scale $\lambda$ describing our probe particle.
Finally we introduce the somewhat non-standard  notation for squares of spacelike vectors: 
\begin{align}
  \chi^2 
  &=
  -\chi_\mu \chi^\mu 
  \ ,
  \qquad
  a^2 
  =
  -a_\mu a^\mu 
  \ ,
  \qquad
  b^2 
  =
  -b_\mu b^\mu
  \ . \label{eqn:squares}
\end{align}
In these equations, one has to carefully distinguish the positive scalars $\chi$, $a$ and $b$ on the left-hand-side versus the contraction of two spatial four vectors on the right-hand-side.

In the probe limit, the second black hole acts as a background and is not dynamical, which in the most general case is
Kerr.
Specifically, we work in Kerr-Schild coordinates with the metric 
\be
  h_{\mu\nu}^{\rm Kerr} = -f k_\mu k_\nu
  \ ,
\ee
with a scalar function $f$ and a four vector $k_\mu$ given by
\begin{align}
f
&=
\frac{2G M r^3}{r^4+(A\cdot x)^2}
\ , 
\\
k_\mu
&=
\eta_\mn\Big(
  V^\nu
  -
  \frac{r}{r^2+A^2}n^\nu
  +\frac{(A\cdot x)}{r}A^\nu
  \Big)
  +
  \frac{\varepsilon_{\mu\nu\rho\kappa} V^{\nu}A^{\rho}x^{\kappa}}{r^2+A^2} 
  \ .
\end{align}
Here $A^\mu$ is the Pauli-Lubanski vector of the background Kerr black hole and the scalar $r$ is implicitly defined through
\be
-\frac{(n\cdot n)}{(r^2+A^2)}+\frac{(A\cdot x)^2}{A^2r^2}=1
\ .
\ee
Note that, just as in \eqn{eqn:squares}, we have $A^{2}=-A^{\mu}A_{\mu}$ yet $A\cdot x= x^{\mu}A_{\mu}$.
Furthermore, the vector $n^\mu$ is defined by,
\begin{equation}
  n^\nu=x^\nu-V^\nu(V\cdot x)+A^\nu(A\cdot x)/A^2
  \ .
\end{equation}
This Lorentz-covariant notation for the Kerr metric is useful because it highlights the physical variables of the background black hole, $V^\mu$ and $A^\mu$, and is easily incorporated into the WQFT framework.
As above, the scalar $A$ is defined through the equation
\begin{align}
  A^2:=-A^\mu A_\mu
  \ .
\end{align}
Note, again, that we use the mostly minus metric.

In the COM frame, where $x^\mu=(x^0,\mathbf x)=(t,x,y,z)$, $A^\mu=(0,0,0,A)$, and $V^\mu=(1,0,0,0)$, one explicitly finds the components
\begin{equation}
  k_\mu=\left(1,\;\;\frac{rx+Ay}{r^2+A^2},\;\;\frac{ry-Ax}{r^2+A^2},\;\;\frac{z}{r}\right)
  \ .
\end{equation}

\begin{table}
  \centering
  \renewcommand{\arraystretch}{1.4}
  \begin{tabular}{|m{4em}|m{7.1em}|m{9.3em}|m{11.2em}|}
    \hline
    \cellcolor{lightgray} \hfil $\mathcal{L}_{(R^1S^s,n)}$
    &
    \cellcolor{lightgray} \hfil $s=2$
    &
    \cellcolor{lightgray} \hfil $s=3$
    &
    \cellcolor{lightgray} \hfil $s=4$
    \\
    \hline 
    \cellcolor{lightgray} \hfil $n=1$ &
    \raggedleft 
    $
    \sfrac{1}{m^2} (S\!\cdot\! S)^{\mu\alpha} R_{\mu\dot z\alpha\dot z}$ & 
    \raggedleft 
    $
    \sfrac{1}{m^3} S^\mn (S \!\cdot\! S)^{\sigma\alpha} R_{\mn\alpha\dot z;\sigma}  $ &
    \raggedleft\arraybackslash 
    $
    \sfrac{1}{m^4} (S\!\cdot\! S)^{\rho\sigma} (S\!\cdot\! S)^{\mu\alpha}  R_{\mu\dot z\alpha\dot z;\rho\sigma}$ 
    \\ 
    \hline
    \cellcolor{lightgray} \hfil $n=2$ &
    \raggedleft 
    $
    \sfrac{1}{m^2}S^\mn S^\ab R_{\mu\nu\alpha\beta}$ &
    \raggedleft 
    $
    \sfrac{1}{m^3} S^{\dot z\sigma} S^\mn S^\ab R_{\mu\nu\alpha\beta;\sigma}$ &
    \raggedleft\arraybackslash 
    $
    \sfrac{1}{m^4} (S\!\cdot\! S)^{\rho\sigma} S^\mn S^\ab R_{\mu\nu\alpha\beta;\rho\sigma}$     \\ 
    \hline
  \end{tabular}
  \caption{
    Linear-in-curvature couplings $\mathcal{L}_{(R^1S^s,n)}$ up to quartic order in spin.
    Indices following a semicolon indicate covariant derivatives.
    Here we use Schoonschip notation such that e.g. $R_{\mn\alpha\dot z}=R_{\mn\ab}\dot z^\beta$ and $S^{\dot z\nu}=g_\ab \dot z^\alpha S^{\beta\nu}$.
    Also, we have used the notation $(S\!\cdot\! S)^\mn=g_\ab S^{\mu\alpha} S^{\beta\nu}$.
    Couplings at higher orders in spin may easily be constructed.
  }
\label{table:LinearInCurvatureCouplings}
\end{table} 

\begin{table}[t]
  \centering
  \renewcommand{\arraystretch}{1.4}
  \begin{tabular}{|m{4em}|m{5.8em}|m{9.5em}|m{12.5em}|}
    \hline
    \cellcolor{lightgray} $\mathcal{L}_{(R^2S^s,n)}$ &
    \cellcolor{lightgray}  \hfil $s=0$ &
    \cellcolor{lightgray}  \hfil $s=2$ &
    \cellcolor{lightgray}  \hfil $s=4$ 
    \\ 
    \hline
    \cellcolor{lightgray} \hfil $n=1$ &
    \raggedleft 
    $
    \probeScale^4
    R_{\mu\dot z\nu\dot z}
    R^{\mu\dot z\nu \dot z}
    $ &
    \raggedleft\arraybackslash 
    $
    \sfrac{\probeScale^2}{m^2}
    R_{\mu\dot z\alpha\dot z}
    {R^{\mu}}_{\!\dot z\beta \dot z}
    (S\!\cdot\! S)^{\ab}
    $ &
    \raggedleft\arraybackslash 
    $
    \sfrac1{m^4}
    (R_{\mu\dot z\nu \dot z}
    (S\!\cdot\! S)^{\mu\nu})^2  
    $
    \\
    \hline
    \cellcolor{lightgray} \hfil $n=2$ &
    \raggedleft 
    $
    \probeScale^4
    R_{\mn\ab}
    R^{\mn\ab}
    $ &
    \raggedleft\arraybackslash 
    $
    \sfrac{\probeScale^2}{m^2}
    R_{\mn\alpha \dot z}
    {R^{\mn}}_{\!\beta\dot z}
    (S\!\cdot\! S)^{\ab}
    $ &
    \raggedleft\arraybackslash 
    $
    \sfrac1{m^4}
    R_{\mu \nu \sigma \dot z}
    R_{\alpha \beta\rho \dot z}
    S^{\mu\nu} S^{\alpha\beta} (S\!\cdot\! S)^{\sigma\rho}
    $
    \\
    \hline
    \cellcolor{lightgray} \hfil $n=3$ &
    \raggedleft 
    &
    \raggedleft\arraybackslash 
    $
    \sfrac{\probeScale^2}{m^2}
    R_{\mu\dot z\nu\sigma}
    R^{\mu\dot z\nu \dot z}
    (S\!\cdot\! S)^{\dot z\sigma}
    $ &
    \raggedleft\arraybackslash 
    $
    \sfrac1{m^4}
    (R_{\mn\ab}S^\mn S^\ab)^2  
    $
    \\
    \hline
    \cellcolor{lightgray} \hfil $n=4$ &
    \raggedleft 
    &
    \raggedleft\arraybackslash 
    $
    \sfrac{\probeScale^2}{m^2}
    R_{\mu\nu\ab}
    {R^{\mu\nu}}_{\!\sigma\dot z}S^{\dot z\sigma}S^{\ab}
    $ &
    \raggedleft\arraybackslash 
    $
    \sfrac1{m^4}
    R_{\mu \dot z\nu\dot z}
    {R^\mu}_{\!\ab\dot z} (S\!\cdot\! S)^{\dot z\nu} (S\!\cdot\! S)^{\ab}    $
    \\
    \hline
  \end{tabular}
  \caption{
    Quadratic-in-curvature couplings $\mathcal{L}_{(R^2S^s,n)}$ with columns and rows corresponding to specific values of $s\in\{0,2,4\}$ and $n\in\{1,2,3,4\}$ respectively.
    These couplings are accurate at leading order in $\lambda$.
    As in table~\ref{table:LinearInCurvatureCouplings}, we use Schoonschip notation and indices are raised and lowered with the full metric $g_\mn$.
  }
\label{table:QuadraticInCurvatureCouplings}
\end{table}

\subsection{Non-minimal terms}\label{sec:nonminimalterms}
Non-minimal interactions of the compact body generally appear as couplings between its dynamical variables and the gravitational curvature.
At the leading linear order in curvature ($R^1$), these Wilson couplings are induced solely by the spin degrees of freedom.
At subleading order in curvature ($R^2$) where tidal interactions appear, spinless non-minimal couplings also exist.
The linear-in-curvature couplings are well understood~\cite{Levi:2015msa,Haddad:2024ebn} and only one free coupling exists at each order in spins (starting at quadratic order in spins).
The general structure of quadratic-in-curvature interactions (known as tidal interactions) is much richer in complexity and under current study~\cite{Bautista:2022wjf,Saketh:2023bul}.

In the present work, we categorise the effective couplings as a power series in the length scale $\lambda$.
Generally, due to the object being compact, we will assume the length scale to be approximately the Schwarzschild radius of the probe $\lambda\sim G m$.
However, we will keep $\lambda$ generic as we want to consider the $m\to0$ limit for fixed $\lambda$ which is the (formal) 0SF limit to be considered in this work.
For Kerr black holes, if we put $\lambda\to Gm$, the dimensionless spin length is limited by $\chi\le1$.

We write the non-minimal part of our action as
\begin{align}\label{eq:nmAction}
  S_{\rm nm}
  =
  m 
  \int \d\tau
  \sum_A \wilson_{A}(\chi^2) 
  \mathcal{L}_A
  \ ,
\end{align}
 where the sum runs over different non-minimal interactions labelled by $A=(R^rS^s,n)$.
Here $r$ counts orders in $R_{\mn\ab}(x)$, $s$ counts orders in $S^\mn(\tau)$ and $n$ labels different operators at the same order.
The interaction terms are $\mathcal{L}_A$ with corresponding (Wilsonian) couplings  $C_A$.
We normalise the interactions such that $C_A$, and therefore $\cL_A$, are dimensionless.
The couplings $C_A$ may generally depend on the (conserved) spin length $\chi^2$ as indicated in \eqn{eq:nmAction}.

All dependence of the mass $m$ in the interactions $\cL_A$ can be exchanged in favour of $\lambda\sim Gm$.
That is, we allow dependence on the worldline variables $p^\mu$ and $S^\mn$ only through $p^\mu/m\sim\dot z^\mu$ and $S^\mn/m\sim \lambda\chi$.
This implies that we have a non-trivial $m\to0$ limit when we keep $\lambda$ finite.
The variable $\lambda$ controls our EFT expansion and higher-order interactions are suppressed by higher powers of $\lambda$.
In this work, we will consider all interactions to quartic order in $\lambda$.
This gives rise to $R^1$ interactions at orders $\lambda^2$, $\lambda^3$ and $\lambda^4$ and $R^2$ interactions at $\lambda^4$ with all operators at $R^3$ and beyond suppressed by higher powers of $\lambda$.

Commonly, in the EFT of spins, one changes basis from the curvature tensor $R_{\mn\ab}(x)$ and spin tensor $S^\mn(\tau)$ to variables that have simpler transformation properties under the SO(3) symmetries of the local frame of the point particle: Namely the electric and magnetic curvature and the spin vector and mass moment.
In this work, however, it is advantageous to use a basis directly in terms of the spin tensor and Riemann curvature tensor.
This avoids the introduction of the Levi-Civita symbol and results in interactions more easily expressed in terms of the bosonic oscillators, $\alpha^{\mu}$ and $\bar\alpha^{\mu}$. 

Importantly, we want to impose a spin supplementary condition (SSC) which constrains half of the degrees of freedom of the spin tensor $S^\mn$.
Thus, with an SSC, one propagates only three degrees of freedom corresponding to a spatial spin vector.
In particular, we choose to enforce the covariant SSC which implies that $S^\mn \dot x_\nu=0$ should be conserved upon using the equations of motion.
In terms of the bosonic oscillators, this translates to the constraint $\alpha^\mu \dot x_\mu =0$.
Conservation of the SSC on the equations of motion is ensured at the level of the action by adding couplings designed and tuned to preserve the SSC constraint.\footnote{This approach was also discussed in Ref.~\cite{Haddad:2024ebn} and has similar features as Refs.~\cite{Alaverdian:2025jtw,Bern:2022kto}.}

Our interaction terms, $\cL_{(R^rS^s,n)}$, at order $R^r$ and $S^s$ are given for $r=1$ in Table~\ref{table:LinearInCurvatureCouplings} and for $r=2$ in Table~\ref{table:QuadraticInCurvatureCouplings}.
As noted, we include all possible couplings to fourth order in the probe scale $\lambda$.
At $R^1$ this implies that we have three free couplings, namely $\cL_{(R^1S^s,1)}$ for $s=2,3,4$.
Further, at order $R^2$ we have six free couplings, namely $\cL_{(R^2S^s,n)}$ for $n=1,2$ and $s=0,2,4$.
Further interactions given in the tables are required to conserve the covariant SSC and appear in the action with fixed, \textit{not free}, coefficients.

Using the couplings of tables~\ref{table:LinearInCurvatureCouplings} and~\ref{table:QuadraticInCurvatureCouplings}, our action has the following explicit expansion
\begin{align}
  \label{eq:NonMinimalAction}
  S_{\rm nm}
  \!=\!
  m\!\int\! \d\tau 
  \bigg[
    &\sum_{s=2}^4
    \wilson_{(R^1S^s,1)} \mathcal{L}_{(R^1S^s,1)}
    +\!\!
  \sum_{s\in\{0,2,4\}}\sum_{n=1}^2
  \wilson_{(R^2S^s,n)}
  \mathcal{L}_{(R^2S^s,n)}
  \\
  +
  &\sum_{s=2}^4
  \wilsonZ_{(R^1S^s,2)} \mathcal{L}_{(R^1S^s,2)}
  +\!\!
\sum_{s\in\{2,4\}}\sum_{n=3}^4
\wilsonZ_{(R^2S^s,n)}
\mathcal{L}_{(R^2S^s,n)}
+
  \cO(\probeScale/\Lambda)^5
  \bigg]
  \, .
  \nonumber
\end{align}
Here we have indicated our neglecting of terms of order $\lambda^5/\Lambda^5$ with $\Lambda$ being an external large length scale (compared with $\lambda$).
Also, as noted, the Wilson coefficients of the first line $\wilson_A=\wilson_A(\chi^2)$ are free while the ones of the second line $\wilsonZ_A=\wilsonZ_A(\chi^2)$ are fixed functions of $\wilson_A$.

The fixed values of $\wilsonZ_A$ in terms of $\wilson_A$ are determined by computing the change in $\alpha^\mu \dot x_\mu$ and demanding it to vanish~\cite{Haddad:2024ebn}.
At linear order in curvature we find
\begin{subequations}
\begin{align}
  \wilsonZ_{(R^1S^2,2)}
  &=
  \frac18
  \ ,
  \\
  \wilsonZ_{(R^1S^3,2)}
  &=
  -\frac18
  \Big(
    1
    +
    2\wilson_{(R^1S^2,1)}
    -
    4\wilson_{(R^1S^3,1)}
  \Big)
  \ ,
  \\
  \wilsonZ_{(R^1S^4,2)}
  &=
  \frac14 \wilson_{(R^1S^3,1)}  
  \ .
\end{align}
\end{subequations}
At quadratic order in curvature we find
\begin{subequations}
\begin{align}
  \wilsonZ_{(R^2S^2,3)}
  &=
  -4\chi^2
  \wilson_{(R^1S^2,1)}
  \wilson_{(R^1S^3,1)}
  \ ,
  \\[3pt]
  \wilsonZ_{(R^2S^2,4)}
  &=
  -\frac{\chi^2}{2}
  \Big(
    6\big[
      1+\wilson_{(R^1S^2,1)}
    \big]
    \wilson_{(R^1S^3,1)}
    +
    \wilson_{(R^1S^4,1)}
  \Big)
  \ ,
  \\[3pt]
  \wilsonZ_{(R^2S^4,3)}
  &=
  -\frac18
  \wilson_{(R^1S^2,1)}
  \wilson_{(R^1S^3,1)}
  -
  \frac18 
  \wilson_{(R^2S^4,2)}
  \ ,
  \\[3pt]
  \wilsonZ_{(R^2S^4,4)}
  &=
  -4
  \wilson_{(R^1S^2,1)}
  \big[
    \wilson_{(R^1S^2,1)}
    -
    6\wilson_{(R^1S^3,1)}
  \big]
  +
  8
  \wilson_{(R^2S^4,2)}
  \ .
\end{align}
\end{subequations}
In this way, the action from \eqn{eq:NonMinimalAction} depends only on (3+6) free couplings $\wilson_A$.

Finally, we may also relate our (free) linear-in-curvature coefficients $C_{(R^1S^s,1)}$ to the ones commonly defined~\cite{Haddad:2024ebn,Levi:2015msa}, namely $C_{ES^s}$ and $C_{BS^s}$.
One finds 
\begin{subequations}\label{eq:wilsoncoeff_EBbasis}
\begin{align}
  \wilson_{(R^1S^2,1)}
  &=
  \frac12
  \big[ 
    C_{ES^2}-1
  \big]
  \ ,
  \\
  \wilson_{(R^1S^3,1)}
  &=
  \frac1{12}
  C_{BS^3}
  \ ,
  \\
  \wilson_{(R^1S^4,1)}
  &=
  \frac1{24}
  \big[
  C_{ES^4}-2C_{BS^3}
  \big]
  \ .
\end{align}
\end{subequations}
For black holes the values of these coefficients are simple: $C_{ES^s}=C_{BS^s}=1$.
Similarly, the $R^2$ coefficients for black holes read
\begin{subequations}\label{eq:wilsoncoeff_R2_Kerr}
\begin{align}
  \wilson_{(R^2S^0,2)}
  &=
  -\frac{\chi^4}{48}
  \ ,
  \\
  \wilson_{(R^2S^2,2)}
  &=
  \frac{\chi^2}{8}
  \ ,
\end{align}
\end{subequations}
and all other coefficients $\wilson_{(R^2S^s,n)}$ vanishing.
In addition, for black holes one must insert $\lambda=G m$.

\subsection{Feynman Rules} \label{sect:props}
The split of the metric $g_{\mu\nu}(x)=\eta_{\mu\nu}+h_{\mu\nu}(x)$ induces  a free and an interacting
part of the worldline action $S_{\rm WQFT}$
\begin{align}
S_{\text{free}}& = - m \int d\tau \,
\eta_{\mu\nu}
\Bigl [ 
\sfrac12 \dot z^{\mu}\dot z^{\nu} + \i 
\bar \alpha^{\prime\mu} \dot \alpha^{\prime \nu} 
\Bigr ]\\
S_{\text{int}}& 
\begin{aligned}[t]
=
-m\int d\tau \Bigl [ &
h_{\mu\nu}[x(\tau)]
\big(
\sfrac12 \dot x^{\mu}\dot x^{\nu}
+
\i
\bar\alpha^\mu\alpha^\nu
\big)
+
\i \bar \alpha_{\mu} 
\Gamma^{\mu}{}_{\nu\rho}[x(\tau)]\dot x^{\nu} \alpha^{\rho}+\ldots 
\Bigr ] 
\Bigr |_{\substack{
{x=b+v\tau}+ z\\ \alpha =\alpha_{m}+ \alpha'}}
\end{aligned}\label{eq:Sint}
\end{align}
Note that the interacting part $S_{\text{int}}$ contains terms of order $n\geq 1$ in $z^\mu(\tau)$, through the dependence of $h_{\mu\nu}[x(\tau)]$ on $z$. An important subtlety concerns the fact that in $S_{\rm free}$ only the 
quadratic deflections  appear and we have dropped the linear ones, which evaluate to
boundary contributions at $\tau=+\infty$. This is related to the fact that the path integral
\eqn{eq:pathintegral_definition} actually cuts things short: We need to consider the in-in or Schwinger-Keldysh
formalism here \cite{Jakobsen:2022psy,Kalin:2022hph}, which entails 
a doubling of fields propagating forward and backward in time
and their matching at the $\tau=+\infty$ boundary. As long as we are interested in tree-level one-point functions, the
upshot of this procedure is that (i) the boundary terms in $S_{\rm free}$ cancel, (ii) the propagators are of
retarded type and define a flow of causality, (iii) the relevant $(n+1)$-leg vertices have one outgoing and $n$ ingoing 
legs of causality flow, they are in functional form identical to the ones derived from $S_{\rm int}$ in the (usual)
in-out (single quantum field) structure of \eqn{eq:Sint}, see  \cite{Jakobsen:2022psy} for the derivation.

For the worldline deflections we have the propagator in configuration space 
\begin{align}\label{eq:propagator_worldline_inin}
    G_{\text{in-in}}^{\mu\nu}(\tau_1,\tau_2) 
    & = \langle z^{\nu}(\tau_2)z^{\mu}(\tau_1) \rangle=
    \begin{tikzpicture}
    \begin{feynman}
      \vertex[small, dot, label=above:$z^\mu(\tau_1)$] at (-0.75,0) (a) {};
      \vertex[small, dot, label=above:$z^\nu(\tau_2)$] at (0.75,0) (b) {};
      \vertex[left = 0.5 of a] (w1);
        \vertex[right = 0.5 of b] (w2);
        \draw[fermion,thick] (a)--(b);
        \draw[dotted, thick] (w1)--(a);
        \draw[dotted, thick] (w2)--(b);
    \end{feynman}
  \end{tikzpicture}
    \\&
    =\int_{-\infty}^\infty\frac{\!d\omega}{2\pi}\,e^{-i\omega(\tau_2-\tau_1)}\left(-\frac{\ i\eta^{\mu\nu}}{m}\frac{1}{(\omega+i\epsilon)^2}\right)=i\frac{\ \eta^{\mu\nu}}{m}(\tau_2-\tau_1)\theta(\tau_2-\tau_1)\,,
  \nn
\end{align}
while the spin degrees of freedom propagate as
\begin{subequations}\label{eq:propagators_spin}
\begin{equation}\label{eq:propagator_aab}
  (G^{\alpha\bar\alpha}_{\rm in-in})^{\mu\nu}(\tau_1,\tau_2)=\vev{\bar\alpha^{\prime \nu}(\tau_{2}) \alpha^{\prime \mu}
  (\tau_{1})}=
  \begin{tikzpicture}
    \begin{feynman}
      \vertex[small, dot, label=above:$\alpha^\mu(\tau_1)$] at (-0.75,0) (a) {};
      \vertex[small, dot, label=above:$\bar\alpha^\nu(\tau_2)$] at (0.75,0) (b) {};
      \vertex[left = 0.5 of a] (w1);
        \vertex[right = 0.5 of b] (w2);
        \draw[thick,fermion] (a)--(b);
        \draw[dotted, thick] (w1)--(a);
        \draw[dotted, thick] (w2)--(b);
    \end{feynman}
  \end{tikzpicture}
  =
  \frac{\ \eta^{\mu\nu}}{m}\theta(\tau_2-\tau_1)\,,
  \end{equation}
  \begin{equation}\label{eq:propagator_aba}
    (G^{\bar\alpha\alpha}_{\rm in-in})^{\mu\nu}(\tau_1,\tau_2)=\vev{\alpha^{\prime \nu}(\tau_{2}) \bar\alpha^{\prime \mu}
  (\tau_{1})}=
    \begin{tikzpicture}
      \begin{feynman}
        \vertex[small, dot, label=above:$\bar\alpha^\mu(\tau_1)$] at (-0.75,0) (a) {};
        \vertex[small, dot, label=above:$\alpha^\nu(\tau_2)$] at (0.75,0) (b) {};
        \vertex[left = 0.5 of a] (w1);
          \vertex[right = 0.5 of b] (w2);
          \draw[thick,fermion] (a)--(b);
          \draw[dotted, thick] (w1)--(a);
          \draw[dotted, thick] (w2)--(b);
      \end{feynman}
    \end{tikzpicture}
    =
    -\frac{\ \eta^{\mu\nu}}{m}\theta(\tau_2-\tau_1)\,.
    \end{equation}
\end{subequations}
In the diagrams the arrow denotes the flow of causality. 
Note that the arguments of the propagator flow from left to right (in contrast to the notation using bras and kets).
It is very convenient to collect the worldline fields into a composite flavoured vector:
\begin{align}
  \Z^{\mu}_{I}(\omega):=\{z^{\mu}(\omega), \alpha^{\prime \mu}(\omega),
  \bar\alpha^{\prime \mu}(\omega)\}
  \ .
\end{align}
We may then collect the above propagators into a single one as
\begin{align}\label{eq:propagator_matrix}
 \begin{tikzpicture}[baseline={(current bounding box.center)}]
  \begin{feynman}
    \coordinate (in) at (-0.6,0);
    \coordinate (out) at (1.4,0);
    \coordinate (x) at (-.2,0);
    \coordinate (y) at (1.0,0);
    \draw [thick,fermion] (x) -- (y) ;
    \draw [dotted] (in) -- (x);
    \draw [dotted] (y) -- (out);
    \draw [fill] (x) circle (.08) node [above] {$\mu, \tau_{1}$} node [below] {$I$};
    \draw [fill] (y) circle (.08) node [above] {$\nu, \tau_{2}$} node [below] {$J$};
     \end{feynman}
  \end{tikzpicture}=& \, 
  \mathcal{G}_{IJ}^{\mu\nu}(\tau_1,\tau_2) =
  \langle \Z_{J}^{\nu}(\tau_{2}) \Z^{\mu}_{I}(\tau_{1}) \rangle_{0}=\nn\\
  &
 \left (\begin{matrix}
   G_{\text{in-in}}^{\mu\nu}(\tau_1,\tau_2)  & & \\
   & &   (G^{\alpha\bar\alpha}_{\rm in-in})^{\mu\nu}(\tau_1,\tau_2) \\
    &  - (G^{\alpha\bar\alpha}_{\rm in-in})^{\mu\nu}(\tau_1,\tau_2)  & 
 \end{matrix}\right )_{IJ} \, .
    \end{align}

The interacting part of the action $S_{\rm int}$ of \eqn{eq:Sint} gives rise to Feynman vertices of multiplicities
to all orders in $\mathcal Z^\mu_{I}$. These arise from the path integral as $n$'th order variations of the action with respect to $\mathcal Z^\mu_{I}$. 
The vertices now involve all fields $\mathcal Z^\mu_{I} \in \{z^\mu(\tau),\alpha'^\mu(\tau),\bar\alpha'^\mu(\tau)\}$. Diagrammatically they take the form 
\begin{equation}\label{eq:Zcalvert}
  \begin{split}
  \begin{tikzpicture}[baseline=(p.base)]
    \begin{feynman}
      \diagram[horizontal=pstart to p, layered layout]{pstart--[dotted,thick]p[large,empty dot,thick]};
      \vertex[small, dot, label=right:$\mathcal Z^{I_1}_{\sigma_1}(\tau_1)$, right = 1.2 of p] (l1) {};
      \vertex[small,dot,label=3:$\mathcal Z^{I_2}_{\sigma_2}(\tau_2)$] at ($(p)+(25:1.2)$) (l2){};
      \vertex[small,dot,label=45:$\mathcal Z^{I_n}_{\sigma_n}(\tau_n)$] at ($(p)+(60:1.2)$) (l3){};
      \vertex at ($(p)+(44:1)$) {$\ddots$};
      \draw[dotted,thick] (pstart)--(p);
      \draw[solid,thick,zUndirected] (p)--(l1);
      \draw[solid,thick,zUndirected] (p)--(l2);
      \draw[solid,thick,zUndirected] (p)--(l3);
    \end{feynman}
  \end{tikzpicture}
  &= 
  \mathcal{V}^{I_1 I_2\dots I_n}_{\sigma_1 \sigma_2\dots\sigma_n}(\tau_1,\tau_2\dots\tau_n)
   \\
  &=
  i\left.\frac{\delta}{\delta \mathcal Z_{I_1}^{\sigma_1}(\tau_{1})}\cdots \frac{\delta}{\delta \mathcal Z_{I_n}^{\sigma_n}(\tau_{n})}S_\text{int}[x(\tau),\alpha(\tau),\bar\alpha(\tau)]\right|_{\{z,\alpha',\bar\alpha'\}=0}.
\end{split}
\end{equation}
We note that the dotted line merely indicates the background straight-line motion of the probe. The one-point vertex for a scalar in a generic background $g_{\mu\nu}(x)=\eta_{\mu\nu}+h_{\mu\nu}(x)$ reads,
\begin{subequations}
  \begin{align}
    \begin{tikzpicture}[baseline=-\the\dimexpr\fontdimen22\textfont2\relax] 
    \begin{feynman}
      \diagram[horizontal=pstart to p, layered layout]{pstart--[dotted,thick]p[large,empty dot,thick]};
      \vertex[small, dot, label=right:$z_{\sigma_1}(\tau_1)$, right = 1.2 of p] (l1) {};
      \draw[dotted,thick] (pstart)--(p);
      \draw[solid, thick,zUndirected] (p)--(l1);
    \end{feynman}
  \end{tikzpicture}
    &\begin{aligned}[t]
      \,\,&=\,\,-\frac{im}{2} \left(\partial_{\sigma_1} h_{\mu\nu}- 2\partial_\mu h_{\sigma_1\nu}\right)v^\mu v^\nu
      \Big|_{x^\mu=b^\mu+v^\mu\tau_1}
    \end{aligned}
    \\
    \begin{tikzpicture}[baseline=-\the\dimexpr\fontdimen22\textfont2\relax]
    \begin{feynman}
      \diagram[horizontal=pstart to p, layered layout]{pstart--[dotted,thick]p[large,empty dot,thick]};
      \vertex[small, dot, label=right:$z_{\sigma_1}(\tau_1)$, right = 1.2 of p] (l1) {};
      \vertex[small,dot,label=right:$z_{\sigma_2}(\tau_2)$] at ($(p)+(25:1.32)$) (l2){};
      \draw[dotted,thick] (pstart)--(p);
      \draw[solid, thick,zUndirected] (p)--(l1);
      \draw[solid,thick,zUndirected] (p)--(l2);
    \end{feynman}
  \end{tikzpicture}
      &\begin{aligned}[t]
    \,\,=\,\,-\frac{im}{2}\Big[\left(
      \partial_{\sigma_1}\partial_{\sigma_2} h_{\mu\nu}- 2\partial_{\mu}\partial_{\sigma_2}h_{\sigma_1\nu}
      \right)v^\mu v^\nu\delta(\tau_1-\tau_2)\\
    +2\left(\partial_{\sigma_1} h_{\sigma_2\mu}-\partial_\mu h_{\sigma_1\sigma_2}-\partial_{\sigma_2}h_{\sigma_1\mu}\right)v^\mu\delta'(\tau_1-\tau_2)
    \\
    - 2h_{\sigma_1\sigma_2}\delta''(\tau_1-\tau_2)
    \Big]
    \Big|_{x^\mu=b^\mu+v^\mu\tau_1}\,,
    \end{aligned}
\end{align}
\end{subequations}
Although not manifestly apparent from its expression, the two-point function is symmetric under exchange of the two external lines $1\leftrightarrow2$ (here, one has to be careful in dealing 
with the differentiated $\delta(\tau_{1}-\tau_{2})$).

As a side remark, the vertices of eq. \eqref{eq:Zcalvert} arise also in the ``standard'' two-body WQFT 
analysis reflecting two interacting
worldlines through a resummation of WQFT Feynman
diagrams, allowing \emph{no} deflection modes on the large mass $M$ worldline:
Let us look at the spinless sector for compactness, where we have
\begin{equation}\label{eq:vertex_shorthand}
  \begin{tikzpicture}[baseline=(p.base)]
    \begin{feynman}
      \diagram[horizontal=pstart to p, layered layout]{pstart--[dotted,thick]p[large,empty dot,thick]};
      \vertex[small, dot, label=right:$z_{\sigma_1}(\tau_1)$, right = 1.2 of p] (l1) {};
      \vertex[small,dot,label=3:$z_{\sigma_2}(\tau_2)$] at ($(p)+(25:1.2)$) (l2){};
      \vertex[small,dot,label=45:$z_{\sigma_n}(\tau_n)$] at ($(p)+(60:1.2)$) (l3){};
      \vertex at ($(p)+(44:1)$) {$\ddots$};
      \draw[dotted,thick] (pstart)--(p);
      \draw[solid, thick,zUndirected] (p)--(l1);
      \draw[solid,thick,zUndirected] (p)--(l2);
      \draw[solid,thick,zUndirected] (p)--(l3);
    \end{feynman}
  \end{tikzpicture}
  \quad
  =
  \quad
  \vcenter{\hbox{\begin{tikzpicture}
    \begin{feynman}
      \diagram[horizontal=w1 to w3, layered layout]{w1--[dotted,thick]w2[blob]--[dotted,thick]w3};
      \vertex[above = 1 of w2] (p);
      \vertex[left = 1 of p] (pstart);
      \vertex[small,dot, label=right:$z_{\sigma_1}(\tau_1)$, right = 1.2 of p] (l1) {};
      \vertex[small,dot,label=3:$z_{\sigma_2}(\tau_2)$] at ($(p)+(25:1.2)$) (l2){};
      \vertex[small,dot,label=45:$z_{\sigma_n}(\tau_n)$] at ($(p)+(60:1.2)$) (l3){};
      \vertex at ($(p)+(44:1)$) {$\ddots$};
      \draw[dotted,thick] (pstart)--(p);
      \draw[photon, thick] (w2)--(p);
      \draw[solid, thick,zUndirected] (p)--(l1);
      \draw[solid,thick,zUndirected] (p)--(l2);
      \draw[solid,thick,zUndirected] (p)--(l3);
    \end{feynman}
  \end{tikzpicture}}}\,.
 \end{equation}
Here the lower, decorated ``Schwarzschild'' vertex contains contributions to all orders $n\ge1$ in the $G$ expansion. 
It arises through a summation of all graphs that possess no deflection modes on the heavy (lower)
worldline, viz.
\begin{align}
  \langle 
    h_{\mu\nu}\rangle
  &=
  \begin{tikzpicture}[baseline]
    \begin{feynman}
      \diagram[horizontal=w1 to w3, layered layout]{w1--[dotted,thick]w2[blob]--[dotted,thick]w3};
      \vertex[above = 1.25 of w2,anchor=base] (p) {$h_{\mu\nu}$};
      \draw[photon, thick] (w2)--(p);
    \end{feynman}
  \end{tikzpicture}
  \;
  \\
    &=
  \;
  \begin{tikzpicture}[baseline]
    \begin{feynman}
      \draw[dotted,thick] (-.6,0)--(.6,0);
      \vertex at (0,0) (p1);
      \draw[photon] (p1)--++(0,1.25);
    \end{feynman}
  \end{tikzpicture}
  \;
  +
  \frac12
  \;
  \begin{tikzpicture}[baseline]
    \begin{feynman}
      \draw[dotted,thick] (-1,0)--(1,0);
      \vertex[black,dot] at (0,0.6) (inter) ;
      \draw[photon] (-0.6,0)--(inter);
      \draw[photon] (0.6,0)--(inter);
      \draw[photon] (inter)--(0,1.25);
    \end{feynman}
  \end{tikzpicture}
  \;
  +
  \frac16
  \;
  \begin{tikzpicture}[baseline]
    \begin{feynman}
      \draw[dotted,thick] (-1,0)--(1,0);
      \vertex at (0,0.6) (inter);
      \draw[photon] (-0.6,0)--(inter);
      \draw[photon] (0,0)--(inter);
      \draw[photon] (0.6,0)--(inter);
      \draw[photon] (inter)--(0,1.25);
    \end{feynman}
  \end{tikzpicture}
  \;
  +
  \frac12
  \;
  \begin{tikzpicture}[baseline]
    \begin{feynman}
      \draw[dotted,thick] (-1,0)--(1,0);
      \vertex at (0,0.6) (inter);
      \vertex at (-0.6,0) (p1);
      \vertex at (0.6,0) (p2);
      \vertex at (0,1.25) (top);
      \vertex at ($(inter)!0.5!(p1)$) (verthalf);
      \draw[photon] (p1)--(inter);
      \draw[photon] (p2)--(inter);
      \draw[photon] (inter)--(top);
      \draw[photon] (0,0)--(verthalf);
    \end{feynman}
  \end{tikzpicture}
  \;
  +\cdots
  \nn
\end{align}
This observation has a long history \cite{Duff:1973zz} and was proven to all orders recently 
in \cite{Mougiakakos:2024nku}. The vertex in \eqn{eq:Zcalvert} thus resums an infinite number of Feynman diagrams encountered in the ``standard'' WQFT approach. This underlying picture will be useful once we enhance the present analysis
to the higher self-force (SF) orders.

\section{The Geodesic Equation as a Berends-Giele Equation}\label{sec:schwing-dyson}

The key idea for a recursive formula to compute the impulse $\Delta p^{\mu}_{\rm 0SF}$ becomes manifest upon further refining our diagrammatic language.
 For one, we drop the dotted lines, as only the probe is dynamical and this bookkeeping is
 unnecessary.
Similarly the causality flow is evident from the tree-structure.
Hence we replace the diagrammatics according to
\begin{equation}
  \begin{tikzpicture}[baseline=-\the\dimexpr\fontdimen22\textfont2\relax,scale=0.5]
    \begin{feynman}
      \vertex at (0,0) (w1) {};
      \vertex [large, empty dot, thick, right = 1 of w1] (v1) {};
      \vertex [large, empty dot, thick, right = 1 of v1] (v2) {};
      \vertex [large, empty dot, thick, right = 1 of v2] (v3) {};
      \vertex [right = 1 of v3] (w2) {};
      \draw[dotted,thick] (w1)--(v1);
      \draw[solid,thick,fermion] (v2)--(v3);
      \draw[dotted,thick] (v1)--(v2);
      \draw[solid,thick,fermion] (v1) to[out=-90,in=-90](v3);
      \draw[solid,thick,fermion] (v3)--(w2);
    \end{feynman}
  \end{tikzpicture} \quad\rightarrow\quad
  \begin{tikzpicture}[baseline=-\the\dimexpr\fontdimen22\textfont2\relax]
    \begin{feynman}
      \vertex[large, empty dot, thick] at (0,0) (vL) {};
      \vertex[large, empty dot, thick] at ($(vL)+(-180+45:1)$) (l1) {};
      \vertex[large, empty dot, thick] at ($(vL)+(-180-45:1)$) (l2) {};
      \vertex[right = 1 of vL] (w2) {};
      \draw[solid,thick] (vL)--(l1);
      \draw[solid,thick] (vL)--(l2);
      \draw[solid,thick] (vL)--(w2);
    \end{feynman}
  \end{tikzpicture}
\end{equation}

\subsection{Spinless case}

To begin with we consider the spinless case and introduce a single multiplicity diagram that symbolizes the all-order
result for the one-point function $\langle z^{\sigma}(\tau)\rangle$ with the final
propagator amputated,
\begin{equation}\label{eq:Fdef_scalar}
  \left.\langle 
    z^\sigma(\tau)
  \rangle\right|_\text{amputated}
  \equiv
  F^\sigma(\tau)
  =
  \begin{tikzpicture}[scale=0.5]
    \begin{feynman}
      \diagram[horizontal=a to b]{a[large,thick,square dot]--[thick,solid]b[label=above:$z^\sigma(\tau)$]};
    \end{feynman}
  \end{tikzpicture}\,.
\end{equation}
From \eqn{eq:propagator_worldline_inin} it takes the form of a force (up to an overall factor of $\i$)
\begin{equation}\label{eq:F_scalar_ddzform}
F^\sigma(\tau)=-\i m \langle \ddot z^\sigma(\tau)\rangle\,.
\end{equation}

\begin{table}[h!]\centering \setstretch{1.5}
  \begin{tabular}{|l|c|}
    \hline\hline
  \# of vertices & diagrams\\\hline\hline
  1 & \begin{tikzpicture}[baseline=-\the\dimexpr\fontdimen22\textfont2\relax,scale=0.5]
    \begin{feynman}
      \vertex at (0,0) (w1) {};
      \vertex[large, empty dot, thick, right = 1 of w1] (v1) {};
      \vertex[right = 1 of v1] (w2) {};
      \draw[dotted, thick,opacity=0] (w1)--(v1);
      \draw[solid, thick] (v1)--(w2);
    \end{feynman}
  \end{tikzpicture}\\\hline
  2 & \begin{tikzpicture}[baseline=-\the\dimexpr\fontdimen22\textfont2\relax,scale=0.5] 
    \begin{feynman}
      \vertex at (0,0) (w1) {};
      \vertex[large, empty dot, thick, right = 1 of w1] (v1) {};
      \vertex[large, empty dot, thick, right = 1 of v1] (v2) {};
      \vertex[right = 1 of v2] (w2) {};
      \draw[dotted, thick,opacity=0] (w1)--(v1);
      \draw[solid, thick] (v1)--(v2);
      \draw[solid, thick] (v2)--(w2);
    \end{feynman}
  \end{tikzpicture}\\\hline
  3 & \makecell{
  \begin{tikzpicture}[scale=0.5]
    \begin{feynman}
      \vertex at (0,0) (w1) {};
      \vertex[large, empty dot, thick, right = 1 of w1] (v1) {};
      \vertex[large, empty dot, thick, right = 1 of v1] (v2) {};
      \vertex[large, empty dot, thick, right = 1 of v2] (v3) {};
      \vertex[right = 1 of v3] (w2) {};
      \draw[dotted, thick,opacity=0] (w1)--(v1);
      \draw[solid, thick] (v1)--(v2);
      \draw[solid, thick] (v2)--(v3);
      \draw[solid, thick] (v3)--(w2);
    \end{feynman}
  \end{tikzpicture}\quad\quad 
  $\frac{1}{2}\;$  \marginbox{0 0.4em 0 0.4em}{\begin{tikzpicture}[baseline=-\the\dimexpr\fontdimen22\textfont2\relax]
    \begin{feynman}
      \vertex[large, empty dot, thick] at (0,0) (vL) {};
      \vertex[large, empty dot, thick] at ($(vL)+(-180+45:1)$) (l1) {};
      \vertex[large, empty dot, thick] at ($(vL)+(-180-45:1)$) (l2) {};
      \vertex[right = 1 of vL] (w2) {};
      \draw[solid,thick] (vL)--(l1);
      \draw[solid,thick] (vL)--(l2);
      \draw[solid,thick] (vL)--(w2);
    \end{feynman}
  \end{tikzpicture}
  }
  }\\\hline
  4 & \makecell{
    \begin{tikzpicture}[baseline=-\the\dimexpr\fontdimen22\textfont2\relax,scale=0.5] 
      \begin{feynman}
        \vertex at (0,0) (w1) {};
        \vertex[large, empty dot, thick, right = 1 of w1] (v1) {};
        \vertex[large, empty dot, thick, right = 1 of v1] (v2) {};
        \vertex[large, empty dot, thick, right = 1 of v2] (v3) {};
        \vertex[large, empty dot, thick, right = 1 of v3] (v4) {};
        \vertex[right = 1 of v4] (w2) {};
        \draw[dotted, thick, opacity=0] (w1)--(v1);
        \draw[solid, thick] (v1)--(v2);
        \draw[solid, thick] (v2)--(v3);
        \draw[solid, thick] (v3)--(v4);
        \draw[solid, thick] (v4)--(w2);
      \end{feynman}
    \end{tikzpicture}\\
    $\frac{1}{2}$
    \begin{tikzpicture}[baseline=-\the\dimexpr\fontdimen22\textfont2\relax]
      \begin{feynman}
        \vertex[large, empty dot, thick] at (0,0) (v2) {};
        \vertex[large, empty dot, thick, right = 1 of v2] (v1) {};
        \vertex[right = 1 of v1] (w1) {};
        \vertex[large, empty dot, thick] at ($(v2)+(-180+45:1)$) (v3) {};
        \vertex[large, empty dot, thick] at ($(v2)+(-180-45:1)$) (v4) {};
        \draw[solid,thick] (w1)--(v1);
        \draw[solid,thick] (v1)--(v2);
        \draw[solid,thick] (v2)--(v3);
        \draw[solid,thick] (v2)--(v4);
      \end{feynman}
    \end{tikzpicture} \quad\quad\quad
    \begin{tikzpicture}[baseline=-\the\dimexpr\fontdimen22\textfont2\relax] 
      \begin{feynman}
        \vertex[large, empty dot, thick] at (0,0) (w1) {};
        \vertex[large, empty dot, thick] at ($(w1)+(-180+45:1)$) (ls) {};
        \vertex[large, empty dot, thick, left=1 of ls] (l1) {};
        \vertex[large, empty dot, thick] at ($(w1)+(-180-45:1)$) (s) {};
        \vertex[right = 1 of w1] (w2);
        \draw[solid, thick] (w1)--(w2);
        \draw[solid, thick] (w1)--(ls);
        \draw[solid, thick] (ls)--(l1);
        \draw[solid, thick] (w1)--(s);
      \end{feynman}
    \end{tikzpicture}
    \\
    $\frac{1}{3!}$
    \marginbox{0 0.4em 0 0}{
    \begin{tikzpicture}[baseline=-\the\dimexpr\fontdimen22\textfont2\relax]
      \begin{feynman}
      \vertex at (0,0) (w) {};
      \vertex[large, empty dot, thick, left = 1 of w] (vL) {};
      \vertex[large, thick, empty dot] at ($(vL)+(-180+50:1)$) (l1) {};
      \vertex[large, thick, empty dot] at ($(vL)+(-180-50:1)$) (l2) {};
      \vertex[large, thick, empty dot] at ($(vL)+(-180:1)$) (l3) {};
      \draw[solid, thick] (vL)--(l1);
      \draw[solid, thick] (vL)--(l2);
      \draw[solid, thick] (vL)--(l3);
      \draw[solid, thick] (w)--(vL);
      \end{feynman}
    \end{tikzpicture}}
  }\\
  \hline
\end{tabular}
\caption{Diagrams contributing to $\mathcal F^I_\sigma(\tau)=\left<\mathcal Z^I_\sigma(\tau)\right>|_\mathrm{amputated}$, with up to 4 vertices (exact up to 4PM). Symmetry factors have been indicated as prefactors of the relevant diagrams. For a spinning probe, every propagator propagates either $z$, $\alpha'$ or $\bar \alpha'$.}\label{tab:diagrams_upto_4pm}
\end{table}

This object obeys a diagrammatic Berends-Giele type recursive relation~\cite{Jakobsen:2023ndj,Jakobsen:2023oow},
\begin{equation}\label{eq:Berends-giele}
   \begin{tikzpicture}[scale=0.5]
    \begin{feynman}
      \diagram[horizontal=a to b]{a[large,thick,square dot]--[thick,solid]b};
    \end{feynman}
  \end{tikzpicture}
  \;\;=\;\;
  \begin{tikzpicture}[baseline=-\the\dimexpr\fontdimen22\textfont2\relax]
    \begin{feynman}
      \vertex at (0,0) (w) {};
      \vertex[large, empty dot, thick, left = 1 of w] (vL){};
      \draw[solid, thick] (vL)--(w);
    \end{feynman}
  \end{tikzpicture}+\;\;\;
    \begin{tikzpicture}[baseline=-\the\dimexpr\fontdimen22\textfont2\relax]
      \begin{feynman}
      \vertex at (0,0) (w) {};
      \vertex[large, empty dot, thick, left = 1 of w] (vL) {};
      \vertex[large, thick, square dot, left = 1 of vL] (l1) {};
      \draw[solid, thick] (vL)--(l1);
      \draw[solid, thick] (w)--(vL);
      \end{feynman}
    \end{tikzpicture}+\,\frac{1}{2}
    \,\,
    \begin{tikzpicture}[baseline=-\the\dimexpr\fontdimen22\textfont2\relax]
      \begin{feynman}
      \vertex at (0,0) (w) {};
      \vertex[large, empty dot, thick, left = 1 of w] (vL) {};
      \vertex[large, thick, square dot] at ($(vL)+(-180+45:1)$) (l1) {};
      \vertex[large, thick, square dot] at ($(vL)+(-180-45:1)$) (l2) {};
      \draw[solid, thick] (vL)--(l1);
      \draw[solid, thick] (vL)--(l2);
      \draw[solid, thick] (w)--(vL);
      \end{feynman}
    \end{tikzpicture}
    +
    \,\frac{1}{3!}
    \,\,\,\,
    \begin{tikzpicture}[baseline=-\the\dimexpr\fontdimen22\textfont2\relax]
      \begin{feynman}
      \vertex at (0,0) (w) {};
      \vertex[large, empty dot, thick, left = 1 of w] (vL) {};
      \vertex[large, thick, square dot] at ($(vL)+(-180+50:1)$) (l1) {};
      \vertex[large, thick, square dot] at ($(vL)+(-180-50:1)$) (l2) {};
      \vertex[large, thick, square dot] at ($(vL)+(-180:1)$) (l3) {};
      \draw[solid, thick] (vL)--(l1);
      \draw[solid, thick] (vL)--(l2);
      \draw[solid, thick] (vL)--(l3);
      \draw[solid, thick] (w)--(vL);
      \end{feynman}
    \end{tikzpicture}
    +\cdots
\end{equation}
Here the white dots denote the arbitrary multiplicity vertices $\mathcal{V}^{I_1 \dots I_n}_{\sigma_1\dots\sigma_n}(\tau_1\dots\tau_n)$ of \eqn{eq:Zcalvert}. As each vertex -- black or white -- scales as
$\cO(G)$, this relation may be solved recursively in a PM expansion. Executing this recursion
straightforwardly recovers the result quoted up to 4PM order in Table~\ref{tab:diagrams_upto_4pm}. In appendix \ref{app:sols_eq_of_motion} we show explicitly that $F^\sigma(\tau)$ obeys the geodesic equations.

We now need to translate the diagrammatic recursion \eqn{eq:Berends-giele} to an explicit mathematical expression. For this we introduce the \emph{force functional} $\mathcal F_{\sigma}[z(\tau)]$ given by the force $F_{\sigma}$ for an \emph{arbitrary} trajectory (one which does not
obey the equations of motion) 
\be\label{eq:F1scalar}
\mathcal F_{\sigma}[z(\tau)]\equiv i \frac{\delta}{\delta z^{\sigma}(\tau)}S_\text{int}
 = 
-i \frac{m}{2}\left(
\partial_{\sigma} h_{\mu\nu}\, \dot x^\mu \dot x^\nu
-2\frac{d}{d\tau}\left[h_{\sigma\nu}\,\dot x^\nu\right]\right)
\Bigg|_{
x^\sigma=b^\sigma +v^\sigma \tau+z(\tau)}
\, .
\ee
Letting deflection modes go to zero, we recover the first diagram in \eqn{eq:Berends-giele}
which is nothing but the leading order vertex rule of \eqn{eq:Zcalvert}
\begin{equation}\label{eq:F_onshell}
  \mathcal F_\sigma[z(\tau)]\Bigr |_{z=0} = \mathcal{V}_\sigma(\tau)
  =\begin{tikzpicture}[scale=0.5]
    \begin{feynman}
      \diagram[horizontal=a to b]{a[large,thick,empty dot]--[thick,solid]b};
    \end{feynman}
  \end{tikzpicture}\quad.
\end{equation}
The remaining diagrams of \eqn{eq:Berends-giele} may be expressed in terms of $F_{\sigma}(\tau)$ and variations of $\mathcal F_{\sigma}[z(\tau)]$ in $z$, resulting in a very compact exponential structure 
\begin{equation}\label{eq:Fexpvarz}
  \begin{split}
  F_\sigma(\tau)&=
    \left.\exp\left[\int_{\tau_0,\bar\tau_{0}}F_{\rho}(\tau_{0})
    G^{\rho\bar\rho}(\tau_{0},\bar\tau_{0}) \frac{\delta}{\delta z^{\bar\rho}(\bar \tau_{0})}
    \right]\mathcal F_\sigma[z(\tau)]\right|_{z=0}
     \\ &=
    \cV_{\sigma}(\tau)+\left.\sum_{n=1}^\infty\frac{1}{n!}
    \left[ \prod_{j=1}^{n}
      \int_{\tau_j,\bar\tau_j}
      F_{\rho_j}(\tau_j) G^{\rho_j\bar\rho_j}(\tau_j,\bar\tau_j)
    \frac{\delta}{\delta z^{\bar\rho_j}(\bar\tau_j)}\right]\mathcal F_\sigma[z(\tau)]\right|_{z=0}\,,
  \end{split}
\end{equation}
with the retarded propagator $G^{\rho\bar\rho}(\tau_0,\bar\tau_0)$ of \eqref{eq:propagator_worldline_inin}.
This recursive relation is the mathematical expression of the diagrammatic series of \eqn{eq:Berends-giele}.

The exponent in \eqn{eq:Fexpvarz} can be further simplified. Writing out the variations, noting from \eqn{eq:F1scalar} that $\mathcal F$ only depends on $z$, $\dot z$ and $\ddot z$, we have
\begin{align}\label{eq:calc_of_variations}
  \frac{\delta}{\delta z^{\rho}(\tau_0)} & \mathcal F_\sigma[z(\tau)]
  = \\ 
  \bigg(
    \delta(\tau_0-\tau)&\frac{\partial}{\partial z^{\bar\rho}(\tau)} 
    +\frac{d}{d\tau}[\delta(\tau_0-\tau)]\frac{\partial}{\partial \dot z^{\bar\rho}(\tau)}
    +\frac{d^2}{d\tau^2}[\delta(\tau_0-\tau)]\frac{\partial}{\partial \ddot z^{\bar\rho}(\tau)}
    \bigg)\mathcal F_\sigma[z(\tau)] \nn \,.
\end{align}
Inserting this in the exponential above, and integrating by parts, we find the more manageable expression
\begin{equation}\label{eq:Fscalar_partialderiv}
  F_\sigma(\tau)=
      \exp\left[
        \begin{split}
        \left[\int_{\tau_0}F_{\rho}(\tau_0)
      G^{\rho\bar\rho}(\tau_0,\tau)\right ]
      &\frac{\partial}{\partial z^{\bar\rho}(\tau)}\\
    +\frac{d}{d\tau}\left[\int_{\tau_0}F_{\rho}(\tau_0)G^{\rho\bar\rho}(\tau_0,\tau)\right]&\frac{\partial}{\partial \dot z^{\bar\rho}(\tau)}
    \\
    +\frac{d^2}{d^2\tau}\left[\int_{\tau_0}F_{\rho}(\tau_0)G^{\rho\bar\rho}(\tau_0,\tau)\right]&\frac{\partial}{\partial \ddot z^{\bar\rho}(\tau)}
    \end{split}
      \right]\mathcal F_\sigma[z(\tau)]\Big|_{z=0}\,,
\end{equation}
involving only partial derivatives of $z(\tau)$, $\dot z(\tau)$ and $\ddot z(\tau)$. Just a single integral in $\tau_0$ appears in the exponential, which can be evaluated for any given $F_{\rho}(\tau_0)$ before taking the partial derivatives.

The above result of \eqn{eq:Fscalar_partialderiv} is very general: in its derivation we have only assumed  (i) that $F^\sigma$ obeys the Berends-Giele type recursion of \eqn{eq:Berends-giele}, (ii) that $\mathcal F[z(\tau)]$ depends only on $z$, $\dot z$ and $\ddot z$, and (iii) that boundary terms related to integration by parts drop out. 

Finally, note that the exponential takes a simple form once derivatives in $\tau$ are evaluated. With the propagator of \eqn{eq:propagator_worldline_inin}, one finds explicitly
\begin{align}\label{eq:Fscalar_explicit}
    F_\sigma(\tau)
    &= 
    \exp\Bigg[ \int_{\tau_0}F^{\rho}(\tau_0)
    \frac{i}{m}
    \\
    &\times
    \bigg( 
      (\tau-\tau_0)\theta(\tau-\tau_0)
    \frac{\partial}{\partial z^{\rho}}
  +
  \theta(\tau-\tau_0)
\frac{\partial}{\partial \dot z^{\rho}} 
+
\delta(\tau-\tau_0)
    \frac{\partial}{\partial \ddot z^{\rho}}\bigg)
    \Bigg]\mathcal F_\sigma[z(\tau)]\Big|_{z=0}\, .
    \nn
\end{align}
Unsurprisingly, this is equivalent to a Taylor expansion of $F_{\sigma}(\tau)$ in $x(\tau)$ around the background configuration.

\subsection{Spinning probes - multiple dynamical fields}

Fluctuations in probe spin can be introduced exactly like those on the worldline, and using the composite field $\mathcal Z_I^\mu=\{z^\mu,\alpha'^\mu,\bar \alpha'^\mu\}$ the resulting equations look almost identical to the spinless case.
In fact, the considerations to follow apply for completely general collections of fields.

The diagrammatic Berends-Giele recursion of \eqn{eq:Berends-giele} then still applies to spinning probes - all propagators, including external ones,  now either propagate the worldline deflections or its spin degrees of freedom. Consequently, we find the same exponential structure as in 
\eqn{eq:Fexpvarz} suitably upgraded to the multi-field scenario. To do this, we augment $F^\sigma$ and $\mathcal F^\sigma$ with flavour-indices $I\in \{z, \alpha^{\prime },
  \bar\alpha^{\prime }\}$, and write 
  \begin{subequations}
    \begin{equation}
      F_I^\sigma(\tau)\equiv \langle \bar{\mathcal Z}_I^\sigma \rangle|_\text{amputated}=
      \begin{tikzpicture}[scale=0.5]
        \begin{feynman}
          \diagram[horizontal=a to b]{a[large,thick,square dot]--[thick,solid]b[label=above:$\mathcal Z_I^\sigma(\tau)$]};
        \end{feynman}
      \end{tikzpicture}
      \ ,
    \end{equation}
    \begin{equation}\label{eq:Ft_def}
        \mathcal F^I_\sigma[\mathcal Z(\tau)] \equiv i\frac{\delta S_\mathrm{int}}{\delta \mathcal Z^\sigma_I(\tau)}=i \left(\partial_{\mathcal Z_I^\sigma(\tau)} L_\mathrm{int}[\mathcal Z(\tau)]- \frac{d}{d\tau}\left[\partial_{\dot{\mathcal Z}^\sigma_I(\tau)} L_\mathrm{int}[\mathcal Z(\tau)]\right]\right)
        \ ,
    \end{equation}
  \end{subequations}
where $\mathcal F^\sigma_I$ takes a form identical to \eqn{eq:F1scalar}, in terms of the interaction Lagrangian $L_\mathrm{int}$ associated with $S_\mathrm{int}=\int d\tau L_{\rm int}[\mathcal Z(\tau),\dot{\mathcal Z}(\tau)]$, assuming
$L_{\rm int}$ depends at most on first order derivatives of $\mathcal Z^{\sigma}_{I}$.  $F_I^\sigma(\tau)$ is related to the one-point functions as
\begin{equation}
  F^\sigma_I(\tau)=\begin{cases}
    -im\langle \ddot z^\sigma(\tau)\rangle, \quad &I=z\\
   - m\langle \dot{\bar\alpha}^\sigma(\tau) \rangle, \quad &I=\alpha'\\
    \phantom{-}m\langle \dot{\alpha}^\sigma(\tau) \rangle, \quad &I=\bar\alpha' \,.
  \end{cases}
\end{equation}
The Berends-Giele recursion of \eqn{eq:Berends-giele} then simply generalises to
\begin{equation}\label{eq:F_var_general}
  F^I_\sigma(\tau)=\left (
  \exp\left[\int_{\tau_0,\bar\tau_{0}}F^J_{\rho}(\tau_0)
  G_{JK}^{\rho\bar\rho}(\tau_0,\bar\tau_{0}) \frac{\delta}{\delta \mathcal Z_K^{\bar\rho}(\bar\tau_{0})}
  \right]\mathcal F^I_\sigma[\mathcal Z(\tau)]\right )\Big|_{z=\alpha'=\bar\alpha'=0}
  \, .
\end{equation}
analogous to \eqn{eq:Fexpvarz}. As with \eqn{eq:Fscalar_partialderiv}, we may integrate by parts. One finds
\begin{equation}\label{eq:Fspin_partialderiv}
    \begin{aligned}
  F^I_\sigma(\tau)&=
  \left (\exp\left[
  \begin{split}  
  \left[\int_{\tau_0}F^J_{\rho}(\tau_0)
  G_{JK}^{\rho\bar\rho}(\tau_0,\tau) \right ]\frac{\partial}{\partial \mathcal Z_K^{\bar\rho}(\tau)}\\
  +\frac{d}{d\tau}\left[\int_{\tau_0}F^J_{\rho}(\tau_0)
  G_{JK}^{\rho\bar\rho}(\tau_0,\tau)\right] \frac{\partial}{\partial \dot{\mathcal Z}_K^{\bar\rho}(\tau)}\\
  +\frac{d^2}{d\tau^2}\left[\int_{\tau_0}F^J_{\rho}(\tau_0)
  G_{JK}^{\rho\bar\rho}(\tau_0,\tau)\right] \frac{\partial}{\partial \ddot{\mathcal Z}_K^{\bar\rho}(\tau)}
  \end{split}
  \right]\mathcal F^I_\sigma[\mathcal Z(\tau)]\right )
  \Bigg|_{z=\alpha'=\bar\alpha'=0}
  \ ,
    \end{aligned}
\end{equation}
ie. an exponential structure with only partial derivatives of $\mathcal F$, and a single integral $\int_{\tau_0} F^J_{\rho}(\tau_0)G_{JK}^{\rho\bar\rho}(\tau_0,\tau)$ over $\tau_0$. Upon noticing
\begin{equation}
  \mathcal Z_K^{\bar\rho}(\tau)=\int_{\tau_0}F^J_{\rho}(\tau_0)
  G_{JK}^{\rho\bar\rho}(\tau_0,\tau)
  \ ,
\end{equation}
$F^I_\sigma(\tau)$ takes the form of a Taylor expansion of $\mathcal F_\sigma^I[\mathcal Z]$ in  $\mathcal Z_{K}$ around straight line motion $z=\alpha'=\bar\alpha'=0$,
\begin{equation}
F^I_\sigma(\tau)=\left.
    \exp\left[
    \mathcal Z_K^{\rho}(\tau)\frac{\partial}{\partial \mathcal Z_K^{\rho}}
  +\dot{\mathcal Z}_K^{\rho}(\tau)\frac{\partial}{\partial \dot{\mathcal Z}_K^{\rho}}
  +
  \ddot{\mathcal Z}_K^{\rho}(\tau)\frac{\partial}{\partial \ddot{\mathcal Z}_K^{\rho}}
  \right]
  \mathcal F_\sigma^I[\mathcal Z(\tau)]
  \right|_{z=\alpha'=\bar\alpha'=0}
  \ .
\end{equation}
The form of eq. \eqref{eq:Fspin_partialderiv} is however more useful, since it explicitly exposes the recursive structure and readily facilitates the integration-by-parts (IBP) reduction discussed in the next section.
Explicitly inserting retarded propagators in eq. \eqref{eq:Fspin_partialderiv} and evaluating the $\tau$-derivatives, one finds
\begin{equation}\label{eq:Fspin_explicit}
  F^I_\sigma(\tau)=
    \exp\left[
      \scalebox{0.82}{\text{$
      \begin{split}
      \int_{\tau_0}F_z^{\rho}(\tau_0)
    \frac{i}{m}\Big(
      \begin{aligned}[t]
        (\tau-\tau_0)\theta(\tau-\tau_0)
  \frac{\partial}{\partial z^{\rho}}
  +
  \theta(\tau-\tau_0)
\frac{\partial}{\partial \dot z^{\rho}}
+
\delta(\tau-\tau_0)
    \frac{\partial}{\partial \ddot z^{\rho}}\Big)
        \end{aligned}\\
    +\int_{\tau_0} F^\rho_{\bar\alpha}(\tau_0)\frac{1}{m}\left(\theta(\tau-\tau_0)\frac{\partial}{\partial\alpha^\rho}
    +
    \delta(\tau-\tau_0)\frac{\partial}{\partial\dot{\alpha}^\rho}
    \right)\\
   -\int_{\tau_0} F^\rho_{\alpha}(\tau_0)\frac{1}{m}\left(\theta(\tau-\tau_0)\frac{\partial}{\partial\bar\alpha^\rho}
   +
   \delta(\tau-\tau_0)\frac{\partial}{\partial\dot{\bar\alpha}^\rho}
   \right)
   \end{split}
      $}}
    \right]\mathcal F^I_\sigma[\mathcal Z(\tau)]\Big|_\mathrm{bg},
\end{equation}
which generalises eq. \eqref{eq:Fscalar_partialderiv}.
We have used the shorthand $|_\mathrm{bg}=|_{z=\alpha'=\bar\alpha'=0}$.

\section{Integration in the PM expansion}\label{sec:int_PMexp}

Generating $F^\sigma_I$ from \eqn{eq:Fspin_explicit} at $n$th order in the PM expansion produces $(n-1)$-fold iterated integrals.
The main difficulty when computing observables is dealing with this increasing iterated complexity.
Some of this complexity may be reduced by considering only perturbative effects in the background Kerr spin.
In this case, an integration-by-parts (IBP) programme reduces the integrals to a simple set of just \textit{one}, 0-fold iterated master-integral.
Such factorisation of the iterated structure agrees with previous studies of the Schwarzschild geometry \cite{Bjerrum-Bohr:2021wwt}.
A generalization to arbitrary powers in the background Kerr spin would be interesting.

\subsection{General integral structure}
Our established recursion expresses $F_I^\mu(\tau)$ in terms of (products of) iterated integrals. This structure is best illustrated with a couple examples. Labelling vertices numerically and only roughly sketching the structure, consider the diagrams
\begin{subequations}
\begin{align}
  &\hspace{.65cm}\begin{aligned}
    \begin{tikzpicture}[baseline=-\the\dimexpr\fontdimen22\textfont2\relax] 
      \begin{feynman}
        \vertex[large, empty dot, thick,minimum size=12pt] at (0,0) (v1) {\footnotesize 3};
        \vertex[large, empty dot, thick,minimum size=12pt, right = 1 of v1] (v2) {\footnotesize 2};
        \vertex[right = 1 of v2,empty dot,minimum size=12pt, thick] (v3) {\footnotesize 1};
        \vertex[right = 1 of v3,label={[label distance=-4]93:$\tau$}] (wend) {};
        \draw[solid, thick] (v1)--(v2);
        \draw[solid, thick] (v2)--(v3);
        \draw[solid, thick] (v3)--(wend);
      \end{feynman}
    \end{tikzpicture}
    \quad&\sim\quad I[f_1,f_2,f_3](\tau)\,,\\
  \begin{tikzpicture}[baseline=-\the\dimexpr\fontdimen22\textfont2\relax]
    \begin{feynman}
      \vertex[large, empty dot, thick,minimum size=12pt] at (0,0) (vL) {\footnotesize 1};
      \vertex[large, empty dot, thick,minimum size=12pt] at ($(vL)+(-180+45:1)$) (l1) {\footnotesize 2};
      \vertex[large, empty dot, thick,minimum size=12pt] at ($(vL)+(-180-45:1)$) (l2) {\footnotesize 3};
      \vertex[right = 1 of vL,label={[label distance=-4]93:$\tau$}] (w2) {};
      \draw[solid,thick] (vL)--(l1);
      \draw[solid,thick] (vL)--(l2);
      \draw[solid,thick] (vL)--(w2);
    \end{feynman}
  \end{tikzpicture}\quad&\sim\quad 
  I
  \Big[f_1 \cdot I[f_2] \cdot I[f_3]
  \Big](\tau)\,,
\end{aligned}
  \\
  &\begin{tikzpicture}[baseline=-\the\dimexpr\fontdimen22\textfont2\relax]
    \begin{feynman}
      \vertex[large,empty dot, thick, minimum size=12pt] at (0,0) (vR) {\footnotesize 2};
      \vertex[large, empty dot, thick, minimum size=12pt, right = 1 of vR] (l0) {\footnotesize 1};
      \vertex[large, empty dot, thick, minimum size=12pt] at ($(vR)+(-180+45:1)$) (l1) {\footnotesize 3};
      \vertex[large, empty dot, thick, minimum size=12pt] at ($(vR)+(-180-45:1)$) (l21) {\footnotesize 4};
      \vertex[large, empty dot, thick, minimum size=12pt] at ($(l21)+(-180-20:1)$) (l22) {\footnotesize 5};
      \vertex[right = 1 of l0,label={[label distance=-4]93:$\tau$}] (wend) {};
      \draw[solid,thick] (l22)--(l21);
      \draw[solid,thick] (l21)--(vR);
      \draw[solid,thick] (l1)--(vR);
      \draw[solid,thick] (vR)--(l0);
      \draw[solid,thick] (l0)--(wend);
    \end{feynman}
  \end{tikzpicture}
  \quad \sim \quad
  I\Big[
  f_1\cdot
  I \Big[
    f_2\cdot I[f_3]\cdot I[f_4,f_5]
  \Big]
   \Big]\label{eq:F_generalstructure}
  (\tau)\,,
\end{align}  
\end{subequations}
which are relevant for $\geq$3PM and $\geq$5PM observables respectively. Here $f_i$ indicate (generally different) functions of $\tau$, sourced by the $i$'th vertex. The vector-structure is omitted for notational simplicity. We denote iterated integrals with
\be
I[f_1,f_2,\ldots , f_{i}](\tau):=
 \int_{-\infty}^\tau\!d\tau_1\,f_1(\tau_1)
\int_{-\infty}^{\tau_1}\!d\tau_2\, f_2(\tau_2)
  \cdots
  \int_{-\infty}^{\tau_{i-1}}\! d\tau_i\, f_i(\tau_i)
  \ .
\ee
Note the mixture of products and iterated integrals in \eqn{eq:F_generalstructure}. Pictorially, a product of integrals arises from a split of branches, whereas iteration comes from consecutive vertices on a single branch. At $n$'th order in the PM expansion there are up to $n-1$ integrations --- factorized or iterated. 

The structure of the functions $f_i$ depends on that of $\mathcal F_I^\mu$ and its derivatives in $\mathcal Z_I$. For spinning probes in a Kerr space-time, perturbative in both probe and Kerr spins, the functions $f_i$ take the form
\begin{equation}\label{eq:fn_generalstructure}
  f_i(\tau)\sim \sum_{j\in\mathbb N}\frac{\text{poly}[\tau]}{r(\tau)^j}
  \ ,
\end{equation}
with denominators only dependent on $r(\tau)=\sqrt{\mathbf x(\tau)^2}=\sqrt{b^2+v^2\gamma^2\tau^2}$, and numerators with positive polynomial powers in $\tau$. The $\frac{1}{r(\tau)^j}$ pole-structure is a natural consequence of our physical PM expansion, where we expand in both $G$ and spins. The order of the pole, $j\leq n$, will at most equal the PM order $n$ in question. Crucially for the IBP reduction in section \ref{sec:reduce_int_complex}, no other poles in $\tau$ appear in this regime. We also note that, asymptotically, $f_i$ obeys 
\begin{equation}\label{eq:fn_asymptote}
  f_i(\tau)\sim\frac{1}{r(\tau)}\sim \frac{1}{\tau}
  \ , \quad\quad \text{as $\tau\rightarrow\pm\infty$}
  \ .
\end{equation}
This will become important for the partial fractioning in section \ref{sec:partial_fractions}.

\subsection{Reducing integral complexity}\label{sec:reduce_int_complex}
The question now arises how to compute products of iterated integrals like \eqn{eq:F_generalstructure} with $f_i$ of the general form \eqref{eq:fn_generalstructure}. We take a practical approach, and seek to simplify the integrals in a four-step process; (i) introduce a change of integration variable which rationalises integrands, (ii) apply partial fraction identities to reduce integrand complexity, (iii) integrate by parts to obtain master integrals, and (iv) apply poly-logarithm identities which, remarkably, {\it factorise} the iterated integrals.

\subsubsection{Change of variables}
Consider the structure of $f_i$ presented in \eqn{eq:fn_generalstructure}. The only irrational $\tau$ dependence appears through
\begin{equation}
  r(\tau)=\sqrt{b^2+v^2\gamma^2\tau^2}\,.
\end{equation}
Such expressions are cumbersome for later IBP-reduction, and we therefore seek to rationalise $f_i(\tau)$. To this extent introduce a new variable, $u$, defined by
\begin{equation}\label{eq:udef}
  u\equiv \frac{\tau \gamma v}{b}+\sqrt{1+\left(\frac{\tau \gamma v}{b}\right)^2}\,,
\end{equation}
in terms of which $r$ and $\tau$ take rational forms
\begin{equation}\label{eq:r_tau_uexpr}
  r=\frac{b}{2} \left(u+u^{-1}\right), \quad\quad \tau=\frac{b}{2\gamma v}(u-u^{-1})\,.
\end{equation}
We note the similarity of the variable $u$ with $x$ usually introduced in PM loop computations~\cite{Herrmann:2021lqe,Driesse:2024feo} where $\gamma=(x+x^{-1})/2$ and $v=(x^{-1}-x)/2$.

Plugging eq. \eqref{eq:r_tau_uexpr} into \eqref{eq:fn_generalstructure}, $f_i$ becomes rational in $u$.
The same is true for integrands in \eqn{eq:F_generalstructure},
\begin{equation}\label{eq:fn_integral_uspace}
  \int_{-\infty}^{\tau'}\!d\tau\, f_i(\tau)\rightarrow \int_0^{u'}\!du\,J(u) f_i(u)\,,
\end{equation}
where
\begin{equation}
J(u)\equiv \frac{d\tau}{du} = \frac{b}{v\gamma}\frac{u ^2+1}{2 u^2}
\end{equation}
is the associated (rational) Jacobian.

\subsubsection{Partial fraction identities}\label{sec:partial_fractions}
We now seek to reduce the complexity of integrands in eq. \eqref{eq:fn_integral_uspace} by applying simple partial fraction identities. 

With the use of eq. \eqref{eq:r_tau_uexpr}, $J(u)f(u)$ takes the form
\begin{equation}\label{eq:Jf_uexpansion}
  J(u)f(u)\sim J(u)\frac{\tau^{i-1}}{r^j}\sim u^{j-i-1}\frac{(-1+u^2)^{i-1}}{(1+u^2)^{j-1}}\,,
\end{equation}
where $i,j\in \mathbb N$ and $i\leq j\leq n$, ensuring the asymptotic requirements of eq. \eqref{eq:fn_asymptote} are obeyed.
Again, $n$ is the PM order under consideration.
A subsequent partial-fractioning yields
\begin{equation}\label{eq:Jf_partialfractions}
  J(u)f(u)\overset{\text{partial fraction}}{\longrightarrow} c_1 \frac{1}{u}+c_2\frac{1}{(1+u^2)^i}+c_3\frac{u}{(1+u^2)^j}\,,
\end{equation}
where $i,j\in \mathbb N$ are some positive numerator powers of $(1+u^2)$, merely indicating the structure.
The coefficients $c_1,c_2$ are $c_{3}$ are $u$-independent constants.

\subsubsection{Integration-by-parts and regularisation}\label{sec:IBP_and_regularisation}
We can further reduce powers of $(1+u^2)$ in \eqn{eq:Jf_partialfractions} by integrating by parts. The rational structure of $J(u)f_i(u)$ in $u$ makes systematic reduction possible. Consider the elementary integration-by-parts identity
\begin{equation}
  \frac{g(u)}{(1+u^2)^j}=\frac{1}{2(j-1)}
  \left(
    \frac{d}{du}\left[\frac{u g(u)}{(1+u^2)^{j-1}}\right]
  -\frac{g(u)(3-2j)+u\frac{dg(u)}{du}}{(1+u^2)^{j-1}}
  \right) .
\end{equation}
which relates $\frac{1}{(1+u^2)^j}$ to lower powers of $j$.
Here $g(u)$ could be any function; in our case $g(u)=1$ or $g(u)=u$, or $g(u)$ can depend on integrals in $u$.
Repeated application to \eqn{eq:Jf_partialfractions} and eq. \eqref{eq:fn_integral_uspace} thus iteratively lowers powers of $(1+u^2)$. What is left are (iterated) integrals of simple integrand-letters, $l_i(u)\in \{\frac{1}{u},\frac{1}{1+u^2},\frac{u}{1+u^2}\}$,
\begin{equation}\label{eq:integral_structure}
  \begin{gathered}
  \Delta \mathcal{Z}^\sigma_I\sim 
  \int_{0}^\infty\! du\, l_0(u)
  \int_{0}^u\!du_1\,l_1(u_1)
  \int_{0}^{u_1}\!du_2\, l_2(u_2)
  \cdots
  \int_{0}^{u_{\delta-1}}\! du_{\delta}\, l_{\delta}(u_{\delta})
\end{gathered}\,,
\end{equation}
with up to $\delta=n-1$ iterations at $n$'th PM order. 
A priori the IBP reduction can produce three letters $l_i(u)\in \{\frac{1}{u},\frac{1}{1+u^2},\frac{u}{1+u^2}\}$. We observe however, for all applications considered in this paper, that only two letters appear
\begin{equation}
  l_i(u)\in 
  \Big\{\frac{1}{u},\frac{1}{1+u^2}
  \Big\} \,.
\end{equation}
So far we have not found an explanation for this simplification.

We conclude this section by considering the regularisation of integrals. Looking at eq. \eqref{eq:integral_structure}, it is clear that such regularisation is indeed necessary. For instance
$
  \int_0^\infty\!du\,\frac{1}{u}
$
is IR divergent, as the two boundaries $u=0$ and $u=\infty$ correspond to $\tau=\pm\infty$.
Consequently, we introduce cutoffs for both the upper and lower limits of all integrals, i.e.~$[0,\infty]\to [\Lambda_-\,\Lambda_+]$. We then have
\begin{equation}\label{eq:regularisation_integrals}
  \int_0^\infty\!du\,\frac{1}{u}\rightarrow \int_{\Lambda_-}^{\Lambda_+}\!du\,\frac{1}{u}\,.
\end{equation}
The regulators are only important for final evaluation of integrals, and do not affect the application of IBP identities; $\Lambda_\pm$ will not enter in intermediate boundary terms since these cancel when $\Lambda_+\rightarrow\infty$ and $\Lambda_-\rightarrow0$, for asymptotically flat theories. We therefore keep regulators implicit until integrals are explicitly evaluated at the end of section \ref{sec:factor_integrals}.
\subsubsection{Factorisation of integral structure - polylogarithm identities}\label{sec:factor_integrals}
The (products of) integrals associated with observables, \eqn{eq:integral_structure}, are still in iterated form.
At the $n$'th PM order these involve integrals of polylogarithms of order $n-1$. 
 This apparent complexity contrasts previous studies, where only {\it products} of integrals of {\it polynomials} appear at arbitrary high PM orders \cite{Damgaard:2022jem}.

Indeed, the iterated structure is only superficial, and may be factorised upon use of polylog shuffle identities. Here we show how this comes about, and conjecture that it happens for any point particle subject to a generic asymptotically flat potential in the weak field regime.

Our central observation is that letters appear in symmetric fashion in \eqn{eq:integral_structure}. For a general number of iterated integrals, $\delta$, we observe they always appear in the combination
\begin{equation}\label{eq:itterated_integrals_symmetric_general}
  \Delta \mathcal{Z}^\sigma_I\sim 
  \sum_{\text{Sym}[l_i]}\left(\int_{0}^\infty\! du\, l_0(u)
  \int_{0}^u\!du_1\,l_1(u_1)
  \int_{0}^{u_1}\!du_2\, l_2(u_2)
  \cdots
  \int_{0}^{u_{d-1}}\! du_{\delta}\, l_{\delta}(u_{\delta})\right)\,,
\end{equation}
where Sym[$l_i$] denotes all permutations of letters $\{l_i\}$. Computing $\Delta \mathcal Z_I^\sigma$ at $n$'th PM order, the iteration depth $\delta$ is at most equal to $n-1$,
\begin{equation}
  \delta\leq n-1\,.
\end{equation}
Concretely, for $\delta=1$ iterated integrals, which appear at 2PM, 
\begin{equation}
    \Delta \mathcal{Z}^\sigma_I\sim 
    \left(\int_{0}^\infty\! du\, l_0(u)
    \int_{0}^u\!du_1\,l_1(u_1)+\int_{0}^\infty\! du\, l_1(u)
    \int_{0}^u\!du_1\,l_0(u_1)\right),
\end{equation}
where $l_0,l_1\in \{\frac{1}{u},\frac{1}{1+u^2}\}$. 
The symmetric combination of integrals in eq. \eqref{eq:itterated_integrals_symmetric_general} may be explicitly factorised by use of a shuffle identity \cite{Duhr:2019tlz}
\begin{equation}\label{eq:perm_integral_identity}
  \sum_{\text{Sym}[l_i]}\left(\int_{a}^b\! du\, l_0(u)
  \int_{a}^u\!du_1\,l_1(u_1)
  \int_{a}^{u_1}\!du_2\, l_2(u_2)
  \cdots
  \int_{a}^{u_{\delta-1}}\! du_\delta\, l_\delta(u_\delta)\right)
  =
  \prod_{i=0}^\delta \int_a^b\!du\, l_i(u)\,,
\end{equation}
for any functions $l_i(u)$, and $a<b$ are arbitrary integration limits. Here $\text{Sym}[l_i]$ denotes all possible permutations of $\{l_i\}$. A proof is given in Appendix \ref{app:shuffle_identity_factorisation}. Applying eq. \eqref{eq:perm_integral_identity} to eq. \eqref{eq:itterated_integrals_symmetric_general}, $\Delta \mathcal{Z}^\sigma_I$ takes an entirely {\it factorised} form in terms of very simple integrals 
\begin{equation}\label{eq:factorisation_integrals}
  \Delta \mathcal Z^\sigma_I\sim 
  \left (\int_0^\infty\!du\, \frac{1}{1+u^2} \right )^{\delta+1}\,.
\end{equation}
Such a remarkable factorisation is by no means evident from the recursive structure alone; it only becomes apparent after integration-by-parts. Factorisation {\it only} happens at the level of observables
\begin{equation}
  \Delta \mathcal{Z}_I^\sigma\sim \int_{-\infty}^\infty\!d\tau\,F_I^\sigma(\tau)\,,
\end{equation}
and {\it not} on the level of $F^\sigma_I(\tau)$, which continues to have iterated structure.
This highlights the difference in complexity of the on-shell observables in contrast to the off-shell trajectory.

Note that in eq. \eqref{eq:factorisation_integrals} the $\frac{1}{u}$ letter drops out completely. This happens for two reasons, (i) iterated integrals in $\frac{1}{u}$ cancel already existing products of integrals (see eq. \eqref{eq:F_generalstructure}), and (ii) remaining products of $\frac{1}{u}$ integrals cancel when taking limits of the (implicit) $\Lambda_{\pm}$ regulators (see section \ref{sec:IBP_and_regularisation} and 
eq.~\eqref{eq:regularisation_integrals}). Indeed, one expects this cancellation since $\int_{\Lambda_-}^{\Lambda_+}\!du\, \frac{1}{u}$ would lead to divergent observables when $\Lambda_\pm$ are taken to $0$ and $\infty$ respectively.

\section{Probe Observables to seventh physical PM order}\label{sec:observables}
\begin{figure}
  \includegraphics[width=.55\textwidth]{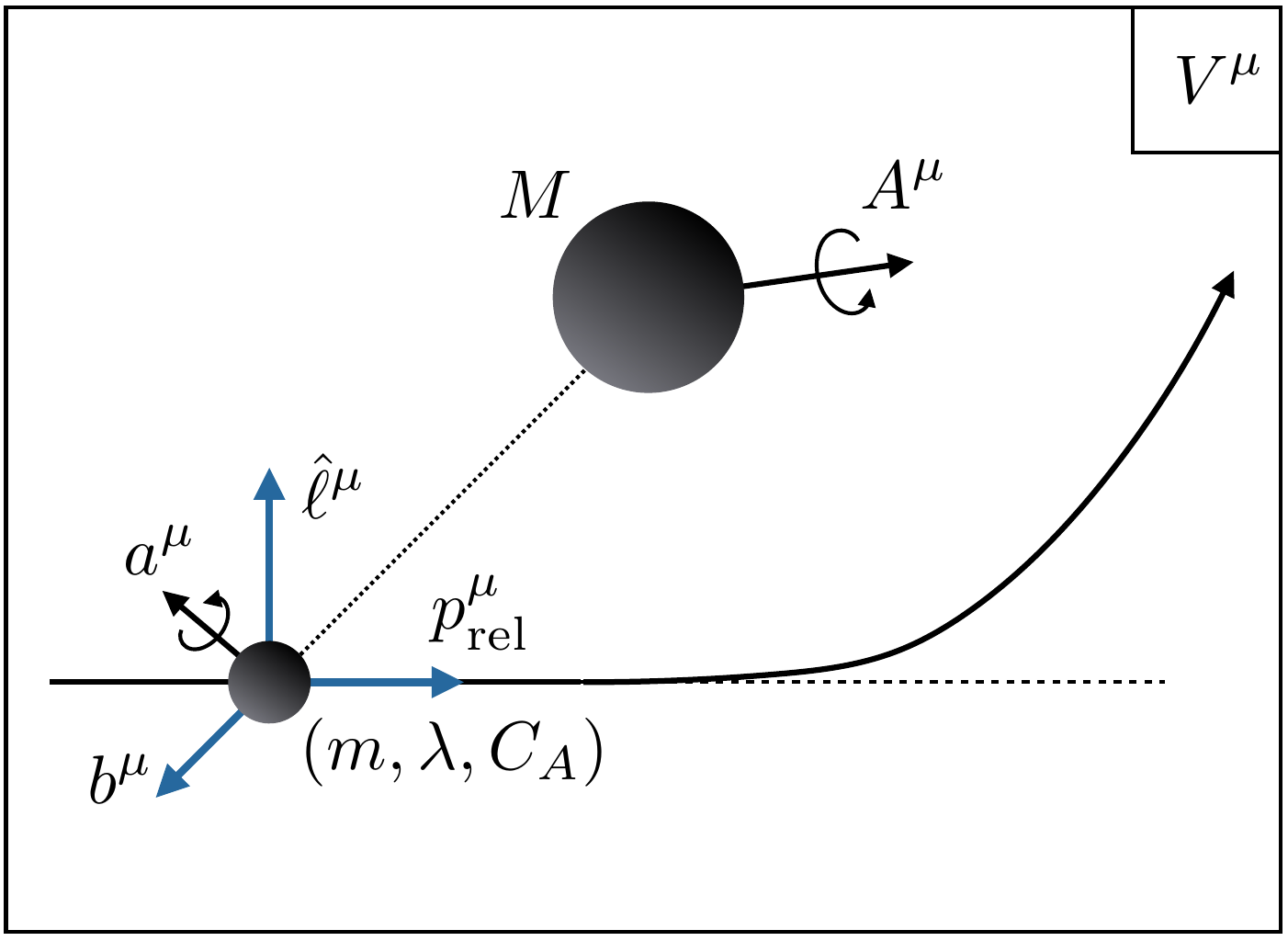}
  \centering
  \caption{
    Kinematics of the two body scattering event in the frame of the background Kerr black hole.
    Each body has mass and spin $(M,A^\mu)$ and $(m,a^\mu)$.
    In addition the probe particle has a length scale $\lambda$ and Wilsonian couplings $C_A$.
    The orthogonal impact parameter is $b^\mu$, the spatial momentum vector of the probe is $p_{\rm rel}^\mu$ and $\hat\ell$ points along the spatial angular momentum of the probe.
  }
  \label{fig:kinematics}
\end{figure}

Finally, we employ our framework to compute scattering observables of the probe in the Kerr background spacetime.
Namely, we compute its change in four-momentum, the impulse $\Delta p^\mu$, and its change in the Pauli-Lubanski vector, the spin kick $\Delta a^\mu$.
Initial state variables were introduced in Sec.~\ref{sec:GFPL} and are summarised in Fig.~\ref{fig:kinematics}.

Each object is characterized by its mass, four velocity and spin vector denoted by $(m,v^\mu,a^\mu)$ and $(M,V^\mu,A^\mu)$ respectively.
Further, the relative impact parameter is $b^\mu=b\,\hat b^\mu$ which points towards the probe.
The relative momentum $p_{\rm rel}^\mu$ is given by the spatial part of the probe momentum,
\begin{align}
  m v^\mu
  =
  m\gamma V^\mu + p_{\rm rel}^\mu
  \ ,
\end{align}
with $|p_{\rm rel}^\mu|=m\gamma v$ and $\hat p_{\rm rel}^\mu=p_{\rm rel}^\mu/|p_{\rm rel}^\mu|$.
Finally, the unit vector $\hat\ell^\mu$ points out of the plane spanned by $\hat b^\mu$ and $\hat p_{\rm rel}^\mu$,
\begin{align}
  \hat \ell^\mu
  =
  \eps^\mu_{\ \nu\rho\sigma}
  \hat b^\nu 
  \hat p_{\rm rel}^\rho
  V^\sigma
  \ ,
\end{align}
which points in the direction of probe angular momentum. Taken together the three vectors $(\hat b^\mu,\hat p_{\rm rel}^\mu,\hat \ell^\mu)$ form a right-handed basis of the spatial subspace to $V^\mu$.
Furthermore, it is convenient to introduce the unit-normalized spin vector,
\begin{align}
  \hat A^\mu = A^\mu/A 
  \ .
\end{align}

\subsection{Schematic form and power counting}
The observables are naturally expanded in the physical PM parameter $GM/b$, the Kerr background spin $A/GM$ and the probe length scale $\lambda/GM$.
It is convenient to include a $1/v^2$ in the PM parameter to have a transparent PN limit (where $b v^2\sim GM$) such that we use $GM/v^2 b$.
In this way, we define expansion coefficients $\Delta v^\mu_{(n,m,l)}$ and $\Delta \chi^\mu_{(n,m,l)}$ via
\begin{subequations}\label{eq:dv_dchi_def}
  \begin{align}
    \frac{\Delta p^\mu}{\gamma v m}
    &=
    \sum_{n=1}^\infty
    \sum_{k=0}^\infty
    \sum_{l=0}^\infty
    \Big(
      \frac{G M}{v^2 b}
    \Big)^{n+k+l}
    \Big(
      \frac{A}{G M}
    \Big)^k
    \Big(
      \frac{\probeScale}{G M}
    \Big)^l
    \Delta v_{(n,k,l)}^{\mu}
    +
    \cO\Big(\frac{m}{M}\Big)
    \ ,
    \\
    \frac{\Delta a^\mu}{\lambda}
    &=
    \sum_{n=1}^\infty
    \sum_{k=0}^\infty
    \sum_{l=1}^\infty
    \Big(
      \frac{G M}{v^2 b}
    \Big)^{n+k+l-1}
    \Big(
      \frac{A}{G M}
    \Big)^k
    \Big(
      \frac{\probeScale}{G M}
    \Big)^{l-1}
    \Delta \chi_{(n,k,l)}^{\mu}
    +
    \cO\Big(\frac{m}{M}\Big)
    \ .
  \end{align}
\end{subequations}
The overall normalizations of the left-hand-sides imply that the expansion coefficients correspond to the changes in the dimensionless variables $v^\mu$ and $\chi^\mu$ which both have a finite, non-trivial $m\to0$ limit (for fixed $\lambda$).

We have computed all contributions to the impulse and spin kick for which $(n+k+l)\le7$ and $l\le4$.
The combination $(n+k+l)=\epsilon_{\rm PM}$ counts the physical PM order and we thus provide probe observables to the seventh PM order ($\eps_{\rm PM}\le7$).
The restriction to $l\le4$ is due to our action being constructed only to the $\lambda^4$ precision.
We note that, for fixed $\lambda$, our results are the leading-order $m\to0$ contributions.
However, for compact bodies, one would assume $\lambda\sim G m$ and thus a physical SF power counting (for compact objects) would assign a physical $l$SF order to the provided impulse.

The dimensionless, vectorial expansion coefficients of the impulse and spin kick, $\Delta v^\mu_{(n,k,l)}$ and $\Delta \chi^\mu_{(n,k,l)}$, may be expressed in terms of the Wilson coefficients $C_A$, the four vectors $\hat b^\mu$, $\hat p_{\rm rel}^\mu$, $\hat\ell^\mu$, $V^\mu$, $\hat A^\mu$ and $\chi^\mu$, and the relative velocity $v$.
All dependence on these variables are polynomial with a few exceptions for $v$ to be discussed sec.~\ref{sec:observables_structure} and appendix~\ref{app:observables_structure}.
The expansion coefficients may be expanded on the basis of $\{V^\mu,\hat b^\mu, \hat p_{\rm rel}^\mu,\hat\ell^\mu\}$ and so, too, may the spin vectors $\hat A^\mu$ and $\chi^\mu$.
The inner product $\chi_\mu V^\mu$ may be eliminated with the SSC $0=v^\mu \chi_\mu=\gamma V^\mu\chi_\mu+\gamma v \hat p_{\rm rel}^\mu \chi_\mu$ and in this manner, we express the observables in terms only of spatial inner products.

\subsection{Checks}
The probe observables satisfy several internal consistency checks.
Namely, for each conserved variable, we may check that its change during scattering vanishes.
The ones involving $a^\mu$, $v^\mu$ and $V^\mu$ are: $a^\mu a_\mu=-\lambda^2\chi^2$, $a^\mu v_\mu=0$, $v^\mu v_\mu=1$ and $V^\mu v_\mu=\gamma$.
Their respective conservation laws read
\begin{subequations}
\begin{align}
  0
  &=
  2 a^\mu\Delta a_\mu + \Delta a^\mu \Delta a_\mu
  \ ,
  \\
  0
  &=
  \Delta a^\mu v_\mu +
  a^\mu \Delta v_\mu 
  +
  \Delta a^\mu \Delta v_\mu
  \ ,
  \\
  0
  &=
  2v^\mu \Delta v_\mu +
  \Delta v^\mu \Delta v_\mu 
  \ ,
  \\
  0
  &=
  V^\mu \Delta v_\mu
  +
  \cO(m/M)
  \ ,
\end{align}
with $\Delta v^\mu=\Delta p^\mu/m$.
Further, at the level of the bosonic oscillators we may verify that the covariant SSC is conserved.
Those checks correspond to the conservation of $\alpha_{m}^\mu v_\mu=\bar\alpha_{m} ^\mu v_\mu=0$: 
\begin{align}
  0
  &=
  \alpha^\mu_{m} \Delta v^\mu 
  +
  \Delta \alpha^\mu v^\mu 
  +
  \Delta \alpha^\mu 
  \Delta v^\mu 
  \ ,
  \\
  0
  &=
  \bar\alpha^\mu_{m} \Delta v^\mu 
  +
  \Delta \bar\alpha^\mu v^\mu 
  +
  \Delta \bar\alpha^\mu 
  \Delta v^\mu 
  \ .
\end{align}
\end{subequations}
It was exactly through these two latter constraints that we determined the values of the extra SSC coefficients $C_A^{\rm SSC}$.

Further, our observables match a variety of results from the literature. The following observables are reproduced;
\begin{itemize}
  \item
  The impulse and spin kick given by Ref.~\cite{Akpinar:2025bkt}, for Kerr values of $\wilson_A$, $k=0$ and all $n\le3$, $l\le4$.
  \item
  The impulse for Kerr values of $\wilson_A$ in Ref.~\cite{Bohnenblust:2024hkw}, for all $n\le2$, $l\le4$ and $k\le 5$.
  \item
  The impulse and spin kick from Ref.~\cite{Jakobsen:2022fcj,Jakobsen:2022zsx}, for all $n\le3$ and $(k+l)\le2$ with generic $\wilson_A$.
  \item
  The impulse and spin kick tabulated in Ref.~\cite{Jakobsen:2023ndj,Jakobsen:2023hig}, for all $n\le4$ and $(k+l)\le1$.
  \item
  The aligned spins impulse found in Ref.~\cite{Damgaard:2022jem}, for Kerr values of $\wilson_A$, at $n\leq 6$, $l\leq 2$ and all $k$.
  \item The $n=1$ impulse, for a Kerr background and arbitrary $\wilson_A$ from Ref.~\cite{Aoude:2021oqj}.
  \item The $l\leq 1$ Kerr impulse and spins-kicks of Ref.~\cite{Gonzo:2024zxo}, truncated at $(n+k+l)\leq 7$.
\end{itemize}

\subsection{PM Observables from the PN Limit}\label{sec:observables_structure}
The expansion coefficients $\Delta v^\mu_{(n,k,l)}$ and $\Delta \chi^\mu_{(n,k,l)}$ have a finite PN limit as $v\to0$, namely their 0PN contribution.
A term scaling as $v^{2m}$ then corresponds to a $m$PN correction.
In this section, we will analyze the structure of the $v$ dependence of the observables.
Interestingly, it turns out that in most cases PM observables can be reconstructed from their PN expansions.

Let us first focus on the $v$ dependence of the impulse through $\Delta v^\mu_{(n,k,l)}$.
Except for a few cases at $\lambda^4$, all $v$ dependence is polynomial, see appendix \ref{app:observables_structure} for details.
We parametrize orders of the velocity by its PN order through $v^{2\epsilon_{\rm PN}}$ where $\epsilon_{\rm PN}$ can be any integer or half-integer.
Remarkably, these powers are bounded from above and below by the following inequality:
\begin{equation}\label{eq:dv_bounds}
  \epsilon_\mathrm{spin}\leq \epsilon_\mathrm{PN}\leq \epsilon_\mathrm{PM}+l/2
  \ ,
\end{equation}
where $\epsilon_{\rm PM}=n+k+l$ counts the physical PM order and $\epsilon_{\rm spin}=k+l$ counts the combined spin order (and, more generally, finite size effects).
It is well-known that spin effects appear at surpressed PN orders which is the statement of the lower bound $\epsilon_{\rm spin}$.
The meaning of this lower bound is further refined when one takes into account that even orders in spin come at integer PN orders while odd orders in spin come at half-integer PN orders.
Thus spinless results start at 0PN order while linear-in-spin results start at 1.5PN order.

Most simply, ignoring finite size effects of the probe with $l=0$, this inequality for the probe impulse reads,
\begin{align}
  \epsilon_{\rm spin}\leq
  \epsilon_{\rm PN}
  \leq
  \epsilon_{\rm PM}
  \ ,
\end{align}
where in this case $\epsilon_{\rm spin}$ counts orders of the background Kerr spin.
Thus in this simplest example, we see that $n$PM probe results are fully determined by their $n$PN approximation.
This is a surprising contrast to the usual, more general, statement that $n$PM results fully determine $(n-1)$PN results~\cite{Vines:2016qwa,Buonanno:2024vkx}.

The velocity dependence of the spin kick as described by its expansion coefficients $\Delta \chi_{(n,k,l)}^\mu$ is a bit more involved.
The $v$ dependence is still polynomial if we allow extra factors of the $\gamma$ factor which we parametrize as $\gamma^l v^{2\epsilon_{\rm PN}}$ for even $l$ or $\gamma^{l-1} v^{2\epsilon_{\rm PN}}$ for odd $l$.
The bounds on $\epsilon_{\rm PN}$ are then very similar to the impulse given by,
\begin{align}\label{eq:dchi_bounds}
  \epsilon_{\rm spin}+1
  \leq
  \epsilon_{\rm PN}
  \leq
  \epsilon_{\rm PM}+l
  \ .
\end{align}
Again, the full PM result may be reconstructed from a $(\epsilon_{\rm PM}+l)$PN approximation.

In appendix \ref{app:observables_structure} we elaborate on details regarding both the structure of $\Delta \chi^\mu_{(n,k,l)}$ and $\Delta v^\mu_{(n,k,l)}$, especially how the $\epsilon_\mathrm{PN}$-bounds depend on the Kerr-spin order $k$, and on $(\chi\cdot \hat p_\mathrm{rel})$ terms. In doing so, the bounds of eqs. \eqref{eq:dv_bounds} and \eqref{eq:dchi_bounds} are slightly refined.

\subsection{Aligned Spins Scattering Angle}
Particularly simple dynamics arises when $\ell^\mu$, $a^\mu$ and $A^\mu$ are all parallel (or antiparallel).
In this case, the spin kick is zero and the impulse is restricted to the initial plane and may be encapsulated in a scattering angle $\theta_\mathrm{aligned}$. It obeys
\begin{equation}
  \sin \theta_\mathrm{aligned}=\hat b_\mu \frac{\Delta p^\mu}{|p_{\rm rel}^\mu|}
  \ .
\end{equation}
We have computed probe contributions, $m\to0$, to $\theta_{\rm aligned}$ to seventh PM order and fourth order in $\lambda$ (namely, the same precision as the impulse).
Due to their lengthy expressions we provide these in the accompanying \zenodo submission.
Below, however, we highlight the relatively simple formal 2PM
contribution ($n\le2$), which is well-known for binary Kerr scattering~\cite{Vines:2017hyw,Aoude:2021oqj,Damgaard:2022jem,Aoude:2023vdk,Gonzo:2024zxo,Kosmopoulos:2021zoq,Chen:2021kxt,Bautista:2023szu,Bern:2022kto}.

At second (formal) PM order the full mass-dependence of the scattering angle, i.e.~the angle for binary scattering, is straightforwardly reconstructed from the probe limit.
Namely, the mass-dependence is simply obtained from the probe limit by adding a factor of the total energy,
\begin{gather}
\begin{aligned}
\theta^\mathrm{1PM}_\mathrm{aligned}
&=
\frac E M
\Big(
\theta^\mathrm{1PM}_\mathrm{aligned}
\Big|_{m\to0}
\Big)
\,,
\\
\theta^\mathrm{2PM}_{\rm aligned}
&=
\frac E M\ 
\Big(
  \theta^\mathrm{2PM}_{\rm aligned}
  \Big|_{m\to0}
\Big)
  \,.
\end{aligned}
\end{gather}
with $E$ given by,
\begin{align}
   E=\sqrt{M^2+m^2+2\gamma Mm}
   \ .
\end{align} 
Below we print the 2PM piece of the scattering angle, highlighting a few important properties.

The 1PM scattering angle, written in terms of arbitrary Wilson coefficients $C_{ES^l}$ and $C_{BS^l}$ in the EB basis, reads
\begin{align}\nonumber
  \theta^\mathrm{1PM}_\mathrm{aligned}
  =
  &\frac{GE}{bv^2}
  \begin{aligned}[t]
    \Big(&
     2 \left(v^2+1\right)
    -\frac{4 A_\ell v}{b}
    +\frac{2 A_\ell^2 \left(v^2+1\right)}{b^2}
    -\frac{4 A_\ell^3 v}{b^3}\\
    &
    +\frac{2 A_\ell^4 \left(v^2+1\right)}{b^4}
    -\frac{4 A_\ell^5 v}{b^5}
    +\frac{2 A_\ell^6 \left(v^2+1\right)}{b^6}
  \Big)
  \end{aligned}
  \\\nonumber
  +&\frac{GE a_\ell}{b^2v^2}
  \begin{aligned}[t]
  \Big(&
    -4 v
    +\frac{4 A_\ell \left(v^2+1\right)}{b}
    -\frac{12 A_\ell^2 v}{b^2}
    +\frac{8 A_\ell^3 \left(v^2+1\right)}{b^3}\\
    &
    -\frac{20 A_\ell^4 v}{b^4}
    +\frac{12 A_\ell^5 \left(v^2+1\right)}{b^5}
  \Big)
  \end{aligned}
  \\\nonumber
  +&\frac{GE a_\ell^2 C_{ES^2}}{b^3v^2}
  \begin{aligned}[t]
  \Big(&
    2 \left(v^2+1\right)
    -\frac{12 A_\ell v}{b}
    +\frac{12 A_\ell^2 \left(v^2+1\right)}{b^2}\\
    &
    -\frac{40 A_\ell^3 v}{b^3}
    +\frac{30 A_\ell^4 \left(v^2+1\right)}{b^4}
  \Big)
  \end{aligned}
  \\ \label{eq:theta_1PM_aligned}
  +&\frac{GE a_\ell^3 C_{BS^3}}{b^4v^2}\left(
    -4 v
    +\frac{8 A_\ell \left(v^2+1\right)}{b}
    -\frac{40 A_\ell^2 v}{b^2}
    +\frac{40 A_\ell^3 \left(v^2+1\right)}{b^3}
  \right)\\\nonumber
  +&\frac{GE a_\ell^4 C_{ES^4}}{b^5v^2}\left(
    2 \left(v^2+1\right)
    -\frac{20 A_\ell v}{b}
    +\frac{30 A_\ell^2 \left(v^2+1\right)}{b^2}
  \right)\\[5pt]\nonumber
  +&\mathcal O(a_\ell^5,(a_\ell+A_\ell)^7)\\\nonumber
  =&\sum_{l=0}^\infty \frac{GE}{b^{l+1}v^2}a_\ell^l C_{(E/B)S^l}V_l(v,A_\ell)+\mathcal O(G^2)\,,
  \end{align}
  where
  \begin{align}
    a_\ell
    =
    -a^\mu \hat\ell_\mu
    \ ,
    \quad\quad
    A_\ell
    =
    -A^\mu \hat \ell_\mu
    \ ,
  \end{align}
  are the directed spin lengths.
  An established, nonetheless remarkable feature \cite{Aoude:2021oqj}, the Wilson coefficients $C_{(E/B)S^l}$ only appear as overall factors at each spin order.
  We show this structure in the last equality, by schematically decomposing the angle in terms of polynomials $V_l(v,A_\ell)$, dependent on the velocity and background Kerr spin. We observe that these are related to each other in a very compact way
  \begin{equation}\label{eq:Vl}
    V_l=\frac{b}{l}\frac{d}{dA_\ell}V_{l-1} \Rightarrow V_l=\frac{b^l}{l!}\frac{d^l}{dA_\ell^l}V_0
    \ .
  \end{equation}
The seed of this recursion, $V_0$, corresponds identically to spinless geodesics in Kerr, which is known to all orders in the Kerr spin $A_\ell$ \cite{Damgaard:2022jem},
  \begin{equation}
    V_0=\frac{2 \left(b^2(1+v^2)-2 A_\ell b v\right)}{b^2-A_\ell^2}
    =
    \frac{(1+v)^2}{1+A_\ell/b}+\frac{(1-v)^2}{1-A_\ell/b}
    \ ,
  \end{equation}
allowing us to resum $V_l(v,A_\ell)$ in Kerr spin,
  \begin{align}
    V_l
    &=(-1)^{l}\frac{(1+v)^2}{(1+A_\ell/b)^{l+1}}+\frac{(1-v)^2}{(1-A_\ell/b)^{l+1}}\,.
  \end{align}\label{eq:512a}
  We conjecture that this expression holds at all orders in probe spin, for arbitrary Wilson coefficients, and thus that the scattering angle takes the general form
  \begin{align}\label{eq:theta_resummeda_1PM}
  \theta^\mathrm{1PM}_\mathrm{aligned}=&\sum_{l=0}^\infty \frac{GE}{b^{l+1}v^2}a_\ell^l C_{(E/B)S^l}\left((-1)^{l}\frac{(1+v)^2}{(1+A_\ell/b)^{l+1}}+\frac{(1-v)^2}{(1-A_\ell/b)^{l+1}}\right)
  \end{align} 
  We also notice from \eqn{eq:theta_1PM_aligned} that even powers of spin couple to $v$ while odd spin powers couple to
  $1+v^2$, which may also be shown from the resummed, conjectured \eqn{eq:512a}.

Specialising to a Kerr probe ($C_{ES^l}=C_{BS^l}=1$), note that eq. \eqref{eq:theta_1PM_aligned} with $V_l$ given by eq. \eqref{eq:Vl} takes the form of a Taylor expansion around the sum of spins, $a_\ell+A_\ell$. We can then resum to all orders in both probe and Kerr spin, recovering
\begin{equation}
  \begin{split}
  \theta^\mathrm{1PM}_\mathrm{aligned}
  &=
  \frac{GE}{bv^2}\sum_{l=0}^\infty\frac{1}{l!}a_\ell^l\frac{d^l}{dA_\ell^l}V_0(A_\ell)
  =
  \frac{GE}{bv^2}V_0(A_\ell+a_\ell)\\
  &=
  \frac{GE}{bv^2}\frac{2 \left(b^2(1+v^2)-2 (A_\ell+a_\ell) b v\right)}{b^2-(A_\ell+a_\ell)^2}\,,
  \end{split}
\end{equation}
which only depends on the sums of spins, as expected from e.g.~ref.~\cite{Vines:2017hyw}. 

\begingroup
\allowdisplaybreaks

Finally, we print the $\mathcal O(G^2)$ aligned scattering angle for $A_\ell=0$ but arbitrary Wilson coefficients. Subtracting the Kerr-probe angle, the difference $\Delta \theta^\mathrm{2PM}\equiv \theta^\mathrm{2PM}-\theta^\mathrm{2PM}_\mathrm{Kerr}$ reads 
\begin{align}
  \Delta \theta_\mathrm{aligned}^\mathrm{2PM}
  =
  &
  \frac{G^2ME}{b^2v^4}
  \bigg[
    a_\ell^2
  \begin{aligned}[t]
    \Big(
    &\frac{3 \pi  \Delta C_{ES^2} \left(5 v^4+32 v^2+8\right)}{16 b^2}\Big)
  \end{aligned}\\[1pt]\nonumber
 +&
 a_\ell^3
 \begin{aligned}[t]
  \Big(-\frac{3 \pi  \Delta C_{ES^2} v \left(11 v^2+4\right)}{4 b^3}-\frac{3 \pi  \Delta C_{BS^3} v \left(v^2+4\right)}{2 b^3}\Big)
   \end{aligned}\\[1pt]\nonumber
  +&
  a_\ell^4
  \begin{aligned}[t]
  \Big(
  &\frac{5 \pi  \left(449 v^6+120 v^4-272 v^2-192\right) \Delta C_{ES^2}}{512 b^4 \left(v^2-1\right)}
+\frac{25 \pi  \left(v^2+6\right) v^2\Delta C_{BS^3}}{16 b^4}\\
+&\frac{15 \pi  \left(13 v^4+44 v^2+8\right) (\Delta C_{ES^2})^2}{128 b^4}
+\frac{5 \pi  \left(-37 v^4+120 v^2+72\right) \Delta C_{ES^4}}{128 b^4}\\
+&\frac{15 \pi  \left(71 v^4-192 v^2+16\right) v^2 \Delta C_{(R^2S^4,2)}}{64 b^4 \left(v^2-1\right)}\\
+&\frac{5 \pi  \left(41 v^4+48 v^2+16\right) v^2 \left(\Delta C_{ES^2}\Delta C_{BS^3}-12\Delta C_{(R^2S^4,1)}\right)}{512 b^4 \left(v^2-1\right)}
\Big)
 \end{aligned}\\[1pt]\nonumber
 +&
a_\ell^2\lambda ^2
\begin{aligned}
    \left(-\frac{75 \pi  \left(v^2-2\right) v^2 \Delta C_{(R^2S^2,2)}}{8 b^4}-\frac{75 \pi  \left(5 v^4+16\right) v^2 \Delta C_{(R^2S^2,1)}}{128 b^4 \left(v^2-1\right)}\right)
 \end{aligned}\\[1pt]\nonumber
 +&
 \lambda ^4
 \begin{aligned}[t]
  \left(-\frac{90 \pi  \left(v^2-1\right) v^2 \Delta C_{(R^2S^0,2)}}{b^4}-\frac{45 \pi  \left(11 v^4+8 v^2+16\right) v^2 \Delta C_{(R^2S^0,1)}}{64 b^4 \left(v^2-1\right)}\right)
 \end{aligned}\\[1pt]
 +&
 \cO(\lambda^5)
 \bigg]
 \,,
 \nn
\end{align}
where
\begin{align}
  \Delta C_A = C_A - C_A^{\rm Kerr}
  \ ,
\end{align}
of which the Kerr-values are defined by eqs. \eqref{eq:wilsoncoeff_EBbasis} and \eqref{eq:wilsoncoeff_R2_Kerr}.

\endgroup

\subsection{Spin kick at linear order in $\lambda$}
Finally, we consider leading order contributions to the spin kick $\Delta a^\mu$.
The spin kick is non-zero only for misaligned spins and thus highlights the generality of our observables.

We remind the reader that we work with the covariant SSC and that $\Delta a^\mu$ denotes the change of the covariant Pauli-Lubanski vector.
For this reason, and in contrast to the canonical kick, $\Delta a^\mu$ has both spatial and timelike components.
At leading (first) order in $\lambda$ and zeroth order in $A$, we find
\begin{align}
  \Delta a^\mu =
  \frac{GM}{bv^2}&\Big[
    -2v^2\hat p_\mathrm{rel}^\mu (a\cdot \hat b)
    +
    4v^2\hat b^\mu (a\cdot \hat p_\mathrm{rel})
    +
    2 v V^\mu (a\cdot \hat b)
  \Big]\\[5pt]\nonumber
  +
  \left(\frac{GM}{bv^2}\right)^2&\Big[
    \begin{aligned}[t]
    &\hat p_\mathrm{rel}^\mu \left(-\frac{3}{4} \pi  v^2 \left(v^2+2\right) (a\cdot \hat b)+2 v^2 \left(3 v^2+1\right) (a\cdot \hat p_\mathrm{rel})\right)\\
  +&\hat b^\mu \left(-2 \left(1-v^2\right) v^2 (a\cdot \hat b)+\frac{3}{4} \pi  \left(3 v^2+2\right) v^2 (a\cdot \hat p_\mathrm{rel})\right)\\
  +&\hat V^\mu \left(2 v \left(1-v^2\right) (a\cdot \hat p_\mathrm{rel})+\frac{3}{2} \pi  v^3 (a\cdot \hat b)\right)\Big]
  \end{aligned}\\[5pt]\nonumber
  +
  \left(\frac{GM}{bv^2}\right)^3&\Big[
    \begin{aligned}[t]
    &\hat p_\mathrm{rel}^\mu \left(-2 v^2 \left(v^4+12 v^2+3\right) (a\cdot \hat b)+3 \pi  v^4 \left(2 v^2+3\right) (a\cdot \hat p_\mathrm{rel})\right)\\
  +&\hat b^\mu \left(\frac{3}{2} \pi  v^6 (a\cdot \hat b)+8 \left(v^2+3\right) v^4 (a\cdot \hat p_\mathrm{rel})\right)\\
  +&\hat V^\mu \left(3 \pi  v^3 \left(1-v^2\right) (a\cdot \hat p_\mathrm{rel})+2 v \left(5 v^4+4 v^2-1\right) (a\cdot \hat b)\right)\Big]
  \end{aligned}\\[5pt]\nonumber
  &
  +\mathcal O(G^4,\lambda^2,A,m)
  \ .
\end{align}
up to fourth order in $G$. Naturally, as with all other results, full expressions are found in the accompanying \zenodo submission.

\section{Conclusions and Outlook}
In this work, we presented a novel framework for computing the classical 
observables associated with the scattering of spinning compact objects in fixed, curved 
background spacetimes using worldline quantum field theory (WQFT). We formulated the geodesic and Mathisson–Papapetrou–Dixon equations as Berends-Giele equations for the one-point functions of the worldline fields yielding the leading order (0SF) result of the
gravitational self force expansion. This reformulation enabled a recursive, diagrammatic computation of classical quantities such as the impulse and spin kick to arbitrary, but fixed, orders in Newton’s constant and spin.

A key technical advance lies in the development of a novel integration-by-parts (IBP) formalism in position space, specifically adapted to worldline quantum field theory. This allowed us to reduce the nested integrals arising in the post-Minkowskian (PM) expansion to a minimal set of master integrals, exhibiting remarkable factorization properties. We provided explicit results for the impulse and spin kick observables up to and including the physical 7PM order, capturing all relevant higher-spin and finite size higher-curvature contributions through quartic order in the probe length scale.
Our results go beyond existing computations in both generality and PM order, offering a systematic and modular approach to incorporating spin-induced finite-size effects in relativistic scattering processes. 

Looking ahead, this work provides the foundation for several promising directions. 
Firstly, extending the analysis to higher finite size and spin effects, i.e.~higher
orders in $\lambda$, is a natural next step. Then, our approach could be applied to arbitrary
space-time backgrounds, exploring further the remarkable factorisation of integrals. Moreover, the recursive Behrends-Giele structure suggests an extension to higher self-force (SF) orders, where the second heavy body becomes dynamical. In this context, combining the WQFT approach with known exact Kerr solutions may yield efficient novel strategies for computing higher-order SF corrections using the toolbox of perturbative quantum field theory. 
Here a first step could be to consider simpler backgrounds such as a gravitational shock
wave and, as in the present work, a combined PM and SF expansion~\cite{Bini:2024icd}.
We believe that working in configuration space could turn out to be crucial to advance. 
Finally, given our high order PM and finite size results, it would be interesting to explore the integrability properties beyond the conservation of the Carter constant at these orders extending the work of \cite{Compere:2023alp}. 
This we leave for future work.

\section*{Code Availability}

All our analytical results for the observables (impulse, spin kick, scattering angle) 
up to and including the (physical) 7PM order as well as the SSC preserving Wilson coefficients
are provided in an accompanying  \zenodo submission. The code to generate these, that can be used to generate even higher orders or work with different metrics, can be made accessible by reasonable request to the authors.

\section*{Acknowledgments}
We thank R.~Gonzo, K.~Haddad, G.~Mogull and R.~Patil for discussions and 
important comments.
This work was supported by the European Union through the 
European Research Council under grant ERC Advanced Grant 101097219 (GraWFTy).
Views and opinions expressed are however those of the authors only and do not necessarily reflect those of the European Union or European Research Council Executive Agency. Neither the European Union nor the granting authority can be held responsible for them.

\appendix

\clearpage

\section{A relative sign at linear order in curvature}
The relation of the linear-in-curvature couplings presented in this work, between $C_{(R^1S^s,1)}$ and the commonly used $E/B$ coefficients $C_{E/BS^s}$, were given in eq.~\eqref{eq:wilsoncoeff_EBbasis}.
It can be derived in two different manners.
First, one may compare observables in terms of $C_{(R^1S^s,1)}$ with ones in terms of $C_{E/BS^s}$ and requiring equality fixes the relationship.
Second, one may work at the level of the action, transforming the Riemann curvature and spin tensors used here to the electric and magnetic curvature tensors and the spin vector. Using the second method, however, we end up with a relative sign on the odd-in-spin couplings as compared with standard literature.
We discuss this small discrepancy here.

Translation of our action may be achieved with the following relation for the (Ricci flat) curvature~\cite{Haddad:2024ebn,Vines:2016unv},
\begin{gather}
\begin{aligned}
   R_{\mu\nu\rho\kappa} =&
   \frac1{|\dot x|^2} \Big(
     - 2E_{\mu[\rho}(g_{\kappa]\nu}|\dot x|^2-2\dot x_{\kappa]} \dot x_{\nu})
     + 2E_{\nu[\rho}(g_{\kappa]\mu}|\dot x|^2-2\dot x_{\kappa]} \dot x_{\mu})
    \\ &\qquad\qquad\qquad\qquad\qquad
        + 2B_{[\mu  }{}^{\lambda} \dot x_{\nu]}\dot x^\alpha\varepsilon_{\alpha\rho \kappa  \lambda} 
        + 2B_{[\rho }{}^{ \lambda} \dot x_{\kappa]}\dot x^\alpha\varepsilon_{\alpha\mu \nu  \lambda} 
    \Big)
     \ ,
\end{aligned}
\end{gather}
and the similar relation for the spin tensor,
\begin{align}
  S^{\mu\nu}
  &=
  \frac1{|\dot x|}
  \Big(
  \varepsilon^{\mu\nu \rho\sigma}\dot x_{\rho}S_{\sigma} + Z^{\mu}\dot x^{\nu} -  Z^{\nu}\dot x^{\mu}
  \Big)
  \ ,
\end{align}
where in the following small analysis, we may ignore the mass moment $Z^\mu$.

Insertion of these relations in our action~\eqref{eq:NonMinimalAction} yields (using also identities satisfied at linear order in curvature~\cite{Levi:2015msa}),
\begin{align}
  S
  &=
  -
  \int \d\tau
  m\eta_\mn
  \bigg(
    \tfrac12
    \dot x^\mu \dot x^\nu
    +
    \i 
    \bar\alpha_\mu
    \frac{\d\alpha^\nu}{\d\tau}
  \bigg)
  \\
  &+\!\!
  \int\!\d\tau\bigg[\!
  -\tfrac12 h_\mn \dot x^\mu \dot x^\nu
  \!+
  \tfrac12 \omega_\mu^{ab}
  \dot x^\mu S_{ab}
  -
  \big(
    C_{(R^1S^2,1)}\!+\!\tfrac12
  \big)
  E_\mn S^\mu S^\nu
  \nn
  \\
  &\qquad\qquad
  +
  2C_{(R^1S^3,1)}
  D_\sigma B_\mn S^\sigma S^\mu S^\nu
  +
  \big(
    C_{(R^1S^3,1)}
    +
    C_{(R^1S^4,1)}
  \big)
  D_\rho D_\sigma
  E_\mn
  S^\rho S^\sigma S^\mu S^\nu
  \bigg]
  \nn
  \\
  &\qquad\qquad+
  \cO(Z,R^2,S^5)
  \ .
  \nn
\end{align}
Here we used the spin connection $\omega^{ab}_\mu=e^a_\nu \pat_\mu e^{\nu b}+e^a_\nu e^{\lambda b}\Gamma^\nu_{\mu\lambda}$ with a local frame $e^a_\nu$ to conform with common notation.
The first line represents kinetic terms and the next terms in square brackets are the interactions terms of $S_{\rm int}$.
Elimination of $C_{(R^1S^s,1)}$ in favor of $C_{E/BS^s}$, using eq.~\eqref{eq:wilsoncoeff_EBbasis}, yields:
\begin{align}
  S_{\rm int}
  &=
  \int\d\tau\bigg[
  -\tfrac12 h_\mn \dot x^\mu \dot x^\nu
  +
  \tfrac12 \omega_\mu^{ab}
  \dot x^\mu S_{ab}
  -
  \tfrac12 C_{ES^2}
  E_\mn S^\mu S^\nu
  +
  \tfrac16 C_{BS^3}
  D_\sigma B_\mn S^\sigma S^\mu S^\nu
  \nn
  \\
  &\hspace{4cm}
  +
  \tfrac1{24} C_{ES^4}
  D_\rho D_\sigma
  E_\mn
  S^\rho S^\sigma S^\mu S^\nu
  \bigg]
  +
  \cO(Z,R^2,S^5)
  \ .
  \label{eq:OurAction}
\end{align}
Comparison of this action with the one of Refs.~\cite{Levi:2015msa,Haddad:2024ebn} show that our third term with $D_\sigma B_\mn$ has a relative sign to the actions found there.
This is puzzling since, as far as we are aware, all conventions here equal the ones used there.

As a small test of the above action, we computed the aligned spins (formal) 1PM scattering angle.
This exercise is straightforward.
First, one may neglect the $\cO(Z)$ terms as we have already done throughout this section.
Second, one may directly insert $S^\mu = (0,0,0,S^3)$ assuming the plane of scattering to be the $x$-$y$-plane.
Third, one may linearize the action in $h_\mn$ and use the leading order 1PM $1/r$ potential.
Fourth, and finally, the relevant equation of motion is simple:
\begin{align}
  \Delta p^\mu=-\int_{-\infty}^{\infty} \d\tau \frac{\pat L_{\rm int}}{\pat x_\mu}+\cO(G^2)
  \ .
\end{align}
That is, one may neglect the second term of the Euler-Lagrange equation.
Here we use $S_{\rm int}=\int \d\tau L_{\rm int}$.
We carried out this small test calculation and we found that the action~\eqref{eq:OurAction} reproduces the known (unambigious) 1PM scattering angle.

\section{From variations to partial derivatives}\label{app:var_to_partder}

Here we seek to retrieve \eqn{eq:Fscalar_partialderiv}, starting from \eqn{eq:Fexpvarz} or equivalently from \eqn{eq:Berends-giele}. Consider the operator in the exponential of \eqn{eq:Fexpvarz},
\begin{equation}\label{eq:caterpillar_diagram}
  \begin{tikzpicture}[baseline=-\the\dimexpr\fontdimen22\textfont2\relax]
    \begin{feynman}
    \vertex at (0,0) (w) {};
    \vertex[left = 1 of w,label={[label distance=-10]0:$\delta$}] (vL) {};
    \vertex[large, thick, square dot, left = 1 of vL] (l1) {};
    \draw[solid, thick] (vL)--(l1);
    \end{feynman}
  \end{tikzpicture}
  \equiv
  \int_{\tau_0,\bar\tau_0}\,
  F_{\rho}(\tau_0) 
  G^{\rho\bar\rho}(\tau_0,\bar\tau_0)
  \frac{\delta}{\delta z^{\bar\rho}(\bar\tau_0)}\ ,
\end{equation}
which is responsible for attaching branches to $\mathcal F[z(\tau)]$. We write a diagrammatic representation of this object, including a $\delta$ label as a reminder that we are dealing with an operator.

In \eqn{eq:Fexpvarz}, this operator acts on $\mathcal F[z(\tau)]$, which only depends on $z$, $\dot z$ and $\ddot z$. Therefore, we can write out the variation $\frac{\delta}{\delta z^{\bar\rho}(\bar\tau_0)}$ from \eqn{eq:calc_of_variations} as
\begin{equation}
  \frac{\delta}{\delta z^{\bar\rho}(\bar\tau_0)}\mathcal F_\sigma(z(\tau))
  =
  \left(
    \delta(\bar\tau_0-\tau)\frac{\partial}{\partial z^{\bar\rho}}
    +\frac{d}{d\tau}[\delta(\bar\tau_0-\tau)]\frac{\partial}{\partial \dot z^{\bar\rho}}
    +\frac{d^2}{d\tau^2}[\delta(\bar\tau_0-\tau)]\frac{\partial}{\partial \ddot z^{\bar\rho}}\right)\mathcal F_\sigma[z(\tau)]\ .
\end{equation}
Stripping away $\mathcal F[z(\tau)]$, this expression is inserted in \eqn{eq:caterpillar_diagram}. We treat each term separately, integrating by parts where convenient,
\begin{equation}\label{eq:zpiece}
  \begin{split}
  \text{($\partial_z$ piece)} &= \int_{\tau_0,\bar\tau_0}\,
  F_{\rho}(\tau_0) 
  G^{\rho\bar\rho}(\tau_0,\bar\tau_0)
  \left[\delta(\bar\tau_0-\tau)\frac{\partial}{\partial z^{\bar\rho}}\right]\\
  &
  =\int_{\tau_0}\,
  F_{\rho}(\tau_0) 
  G^{\rho\bar\rho}(\tau_0,\tau)
  \frac{\partial}{\partial z^{\bar\rho}},
  \end{split}
\end{equation}
\begin{equation}\label{eq:dzpiece}
  \begin{split}
\text{($\partial_{\dot z}$ piece)} =& 
\int_{\tau_0,\bar\tau_0}\,
F_{\rho}(\tau_0) 
  G^{\rho\bar\rho}(\tau_0,\bar\tau_0)
\left[  \frac{d}{d\tau}[\delta(\bar\tau_0-\tau)]\frac{\partial}{\partial \dot z^{\bar\rho}}  \right]\\
=&
\int_{\tau_0,\bar\tau_0}\,
F_{\rho}(\tau_0) 
  G^{\rho\bar\rho}(\tau_0,\bar\tau_0)
\left[  -\frac{d}{d\bar\tau_0}[\delta(\bar\tau_0-\tau)]\frac{\partial}{\partial \dot z^{\bar\rho}}  \right]\\
=&
\int_{\tau_0,\bar\tau_0}\,
F_{\rho}(\tau_0) 
\frac{d}{d\bar\tau_0}\left[G^{\rho\bar\rho}(\tau_0,\bar\tau_0)\right]
\delta(\bar\tau_0-\tau)\frac{\partial}{\partial \dot z^{\bar\rho}}\\
=&
\frac{d}{d\tau}\left[\int_{\tau_0}\,
F_{\rho}(\tau_0) 
  G^{\rho\bar\rho}(\tau_0,\tau)\right]
\frac{\partial}{\partial \dot z^{\bar\rho}},
  \end{split}
\end{equation}
where in the third line we assume the boundary term vanishes,
\begin{equation}
  \left.G^{\rho\bar\rho}(\tau_0,\bar\tau_0)
  \delta(\bar\tau_0-\tau)\right|_{\bar\tau_0=-\infty}^{\bar\tau_0=\infty}
  =
  0.
\end{equation}
The $\partial_{\ddot z}$ piece similarly reads
\begin{equation}\label{eq:ddzpiece}
  \begin{split}
\text{($\partial_{\ddot z}$ piece)}
=&\int_{\tau_0,\bar\tau_0}\,
F_{\rho}(\tau_0) 
  G^{\rho\bar\rho}(\tau_0,\bar\tau_0)
\left[
  \frac{d^2}{d\tau^2}[\delta(\bar\tau_0-\tau)]\frac{\partial}{\partial \ddot z^{\bar\rho}}
\right]\\
=&\int_{\tau_0,\bar\tau_0}\,
F_{\rho}(\tau_0) 
  G^{\rho\bar\rho}(\tau_0,\bar\tau_0)
\left[
  \frac{d^2}{d\bar\tau_0^2}[\delta(\bar\tau_0-\tau)]\frac{\partial}{\partial \ddot z^{\bar\rho}}
\right]\\
=&
\frac{d^2}{d\tau^2}\left[\int_{\tau_0}\,
F_{\rho}(\tau_0) 
  G^{\rho\bar\rho}(\tau_0,\tau)
\right]
  \frac{\partial}{\partial \ddot z^{\bar\rho}}\ ,
  \end{split}
\end{equation}
assuming again that boundary terms vanish,
\begin{equation}
\left.G^{\rho\bar\rho}(\tau_0,\bar\tau_0)\frac{d}{d\bar\tau_0}\delta(\bar\tau_0-\tau)\right|^{\bar\tau_0=\infty}_{\bar\tau_0=-\infty}=0,\quad\left.\frac{d}{d\bar\tau_0}[G^{\rho\bar\rho}(\tau_0,\bar\tau_0)]\delta(\bar\tau_0-\tau)\right|^{\bar\tau_0=\infty}_{\bar\tau_0=-\infty}=0\ .
\end{equation}
Collecting the pieces from eqs. \eqref{eq:zpiece}, \eqref{eq:dzpiece} and \eqref{eq:ddzpiece}, we find
\begin{equation}
  \begin{tikzpicture}[baseline=-\the\dimexpr\fontdimen22\textfont2\relax]
    \begin{feynman}
    \vertex at (0,0) (w) {};
    \vertex[left = 1 of w,label={[label distance=-10]0:$\delta$}] (vL) {};
    \vertex[large, thick, square dot, left = 1 of vL] (l1) {};
    \draw[solid, thick] (vL)--(l1);
    \end{feynman}
  \end{tikzpicture}
  =
  \left(
    \begin{split}
  \int_{\tau_0}\,
  F_{\rho}(\tau_0)
  G^{\rho\bar\rho}(\tau_0,\tau)
  &\frac{\partial}{\partial z^{\bar\rho}}
  \\+
  \frac{d}{d\tau}\left[\int_{\tau_0}\,
F_{\rho}(\tau_0) 
  G^{\rho\bar\rho}(\tau_0,\tau)\right]
&\frac{\partial}{\partial \dot z^{\bar\rho}}
\\+
\frac{d^2}{d\tau^2}\left[\int_{\tau_0}\,
F_{\rho}(\tau_0) 
  G^{\rho\bar\rho}(\tau_0,\tau)
\right]
  &\frac{\partial}{\partial \ddot z^{\bar\rho}}
    \end{split}
\right).
\end{equation}
We may now re-insert this operator into the exponential of \eqn{eq:Fexpvarz}, immediately recovering \eqn{eq:Fscalar_partialderiv},
  \begin{equation}
    F_\sigma(\tau)=
      \exp\left[
        \begin{split}
        \int_{\tau_0}F_{\rho}(\tau_0)
      G^{\rho\bar\rho}(\tau_0,\tau)
      &\frac{\partial}{\partial z^{\bar\rho}}\\
    +\frac{d}{d\tau}\left[\int_{\tau_0}F_{\rho}(\tau_0)G^{\rho\bar\rho}(\tau_0,\tau)\right]&\frac{\partial}{\partial \dot z^{\bar\rho}}
    \\
    +\frac{d^2}{d^2\tau}\left[\int_{\tau_0}F_{\rho}(\tau_0)G^{\rho\bar\rho}(\tau_0,\tau)\right]&\frac{\partial}{\partial \ddot z^{\bar\rho}}
    \end{split}
      \right]\mathcal F_\sigma[z(\tau)]\ .
  \end{equation}
 \section{$\mathcal F_\sigma^z$ as solutions to the equations of motion}\label{app:sols_eq_of_motion}
We show explicitly that $\mathcal F^z_\sigma$ solves the geodesic equations of motion for a non-spinning scalar,
\begin{equation}
  \ddot x^\sigma=-\Gamma^\sigma_{\;\mu\nu}\dot x^\mu\dot x^\nu.
\end{equation}
For simplicity, we drop the superscript $z$ and write $\mathcal F^z_\sigma\rightarrow\mathcal F_\sigma$ \footnote{as in eq. \eqref{eq:Fscalar_partialderiv}}.

Our starting point is eq. \eqref{eq:Fscalar_partialderiv}, and its interpretation as a Taylor expansion,
\begin{equation}
  F_\sigma(\tau)=\left.\exp\left[
    z^\rho(\tau) \frac{\partial}{\partial x^\rho}
    +\dot z^\rho(\tau) \frac{\partial}{\partial \dot x^\rho}
    +\ddot z^\rho(\tau) \frac{\partial}{\partial \ddot x^\rho}
    \right]\mathcal F_\sigma[z]\right|_{bg}.
\end{equation}
where $bg$ refers to straight-line motion, $x^\mu=b^\mu+v^\mu\tau$. From eq. \eqref{eq:F1scalar}, $\mathcal F_\sigma[z]$ reads
\begin{equation}
\mathcal F_\sigma[z]=-i \frac{m}{2}\left(
\partial_{\sigma} h_{\mu\nu} \dot x^\mu \dot x^\nu
-2\frac{d}{d\tau}\left[h_{\sigma\nu}(x)\dot x^\nu\right]\right)
\Bigg|_{
x^\sigma=b^\sigma +v^\sigma \tau+z(\tau)}
\, .
\end{equation} 
which when inserted above allows explicit evaluation of the $\frac{\partial}{\partial \ddot z}$ derivatives in the exponential. Noting that $\mathcal F$ is at most linear in $\ddot z$, we find, after a little rewriting,
\begin{equation}
  F_\sigma(\tau)=im\left.\exp\left[
    z^\rho(\tau) \frac{\partial}{\partial z^\rho}
    +\dot z^\rho(\tau) \frac{\partial}{\partial \dot z^\rho}
    \right]\left(-\frac{\partial_{\sigma} h_{\mu\nu}\dot x^\mu\dot x^\nu}{2}+\partial_{\mu} h_{\sigma\nu}\dot x^\mu\dot x^\nu+h_{\sigma\rho}\ddot z^\rho\right)\right|_{bg}.
\end{equation}
From eq. \eqref{eq:F_scalar_ddzform} we have $F_\sigma(\tau)=-im\ddot z_\sigma$, with indices raised and lowered by $\eta_{\mu\nu}$. Inserting above, factors of $im$ cancel and we find
\begin{equation}\label{eq:ddz_geodesic_prelim}
  \eta_{\sigma\rho}\ddot z^\rho
  =
  \left.\exp\left[
    z^\rho(\tau) \frac{\partial}{\partial z^\rho}
    +\dot z^\rho(\tau) \frac{\partial}{\partial \dot z^\rho}
    \right]\left(\frac{\partial_{\sigma} h_{\mu\nu}\dot x^\mu\dot x^\nu}{2}-\partial_{\mu} h_{\sigma\nu}\dot x^\mu\dot x^\nu-h_{\sigma\rho}\ddot z^\rho\right)\right|_{bg}
\end{equation}
The right-hand-side takes the form of a Taylor-expansion of the term in () brackets, which is only dependent on $z$ and $\dot z$. Symmetrising $\rho$ and $\nu$ indices, the first and second term can be rewritten in terms of the Christoffel symbol,
\begin{equation}
  \begin{split}
  \frac{\partial_{\sigma} h_{\mu\nu}\dot x^\mu\dot x^\nu}{2}-\partial_{\rho} h_{\sigma\nu}\dot x^\rho\dot x^\nu
  &=
  \frac{1}{2}\left(
    \partial_{\sigma} h_{\mu\nu}\dot x^\mu\dot x^\nu
    -
    \partial_{\mu} h_{\sigma\nu}\dot x^\mu\dot x^\nu
    -
    \partial_{\mu} h_{\sigma\nu}\dot x^\nu\dot x^\mu
  \right)\\
  &=-\Gamma_{\sigma\mu\nu}\dot x^\mu\dot x^\nu,
  \end{split}
\end{equation}
with $\Gamma_{\sigma\mu\nu}=g_{\sigma\rho}\Gamma^\rho_{\;\mu\nu}$. Resumming the Taylor expansion in eq. \eqref{eq:ddz_geodesic_prelim}, ie. evaluating $\Gamma_{\sigma\mu\nu}\dot x^\mu\dot x^\nu+h_{\sigma\rho}\ddot z^\rho$ at the full trajectory $x^\mu=b^\mu+v^\mu\tau+z^\mu(\tau)$, we find
\begin{equation}
  \eta_{\sigma\rho}\ddot z^\rho
  =
  -\Gamma_{\sigma\mu\nu}\dot x^\mu\dot x^\nu-h_{\sigma\rho}\ddot z^\rho.
\end{equation}
implicitly evaluated at the full trajectory $x^\mu=b^\mu+v^\mu\tau+z^\mu(\tau)$. Finally, pulling $h_{\sigma\rho}\ddot z^\rho$ to the left-hand-side, we recognise $g_{\sigma\rho}=\eta_{\sigma\rho}+h_{\sigma\rho}$. Lowering index $\sigma$ with the full metric, $g$ then yields
\begin{equation}
  \ddot x^\sigma
  =
  -\Gamma^\sigma_{\;\mu\nu}\dot x^\mu\dot x^\nu.
\end{equation}
which is exactly the geodesic equation, noting that $\ddot z^\rho=\ddot x^\rho$.

\section{Factorisation of integrals - proof of eq. \eqref{eq:perm_integral_identity}}\label{app:shuffle_identity_factorisation}
We provide a short proof of eq. \eqref{eq:perm_integral_identity},
\begin{equation}
  \sum_{\text{Sym}[l_i]}\left(\int_{a}^b\! du_0\, l_0(u)
  \int_{a}^u\!du_1\,l_1(u_1)
  \int_{a}^{u_1}\!du_2\, l_2(u_2)
  \cdots
  \int_{a}^{u_{n-1}}\! du_n\, l_n(u_n)\right)
  =
  \prod_{i=0}^n \int_a^b\!du\, l_i(u).
\end{equation}
Introducing Heaviside functions, $\theta(u_i-u_j)$, we can rewrite the LHS as
\begin{equation}\label{eq:permutations_written_out}
  \begin{gathered}
  \begin{split}
  &\sum_{\text{Sym}[l_i]}\left(\int_{a}^b\! du_0\, l_0(u)
  \int_{a}^u\!du_1\,l_1(u_1)
  \int_{a}^{u_1}\!du_2\, l_2(u_2)
  \cdots
  \int_{a}^{u_{n-1}}\! du_n\, l_n(u_n)\right)\\
  &=\int_a^b\!\prod_{i=0}^n du_i\, l_i(u_i)\sum_{\text{Sym}[0,1,2,...,n]}\theta(u_0-u_1)\theta(u_1-u_2)\cdots \theta(u_{n-1}-u_n)\ ,
  \end{split}\\
  \theta(u_i-u_j)=\begin{cases}
    1, \quad u_i>u_j\\
    0, \quad \text{otherwise}
  \end{cases}.
\end{gathered}
\end{equation}
Each product of Heaviside-functions covers a different section of parameter space $\{u_0,u_1,...u_n\}$, without any overlap,
\begin{equation}
  \theta(u_0-u_1)\theta(u_1-u_2)\cdots \theta(u_{n-1}-u_n)=
  \begin{cases}
  1, \quad\quad u_0\geq u_1\geq u_2\geq u_3\geq...\geq u_n\\
  0, \quad\quad \text{otherwise}
  \end{cases}
\end{equation}
The permutations in eq. \eqref{eq:permutations_written_out} corresponds to all possible orderings of $u_i$ in the inequality above, the collection of which exactly covers all parameter-space. We thus have
\begin{equation}
  \sum_{\text{Sym}[0,1,2,...,n]}\theta(u_0-u_1)\theta(u_1-u_2)\cdots \theta(u_{n-1}-u_n)=1
\end{equation}
which when inserted in eq. \eqref{eq:permutations_written_out} immediately recovers eq. \eqref{eq:perm_integral_identity},
\begin{equation}
  \begin{split}
  \sum_{\text{Sym}[l_i]}\left(\int_{a}^b\! du_0\, l_0(u)
  \int_{a}^u\!du_1\,l_1(u_1)
  \int_{a}^{u_1}\!du_2\, l_2(u_2)
  \cdots
  \int_{a}^{u_{n-1}}\! du_n\, l_n(u_n)\right)=\prod_{i=0}^n\int_a^b\! du_i\, l_i(u_i).
  \end{split}
\end{equation}

\section{Structure  of observables}\label{app:observables_structure}
Here we elaborate on the velocity-structure of $\Delta v^\mu_{(n,k,l)}$ and $\Delta \chi^\mu_{(n,k,l)}$, discussed in section \ref{sec:observables_structure}. By considering also the $(\hat {a}\cdot \hat{p}_\mathrm{rel})$ and Kerr-spin dependence of expressions, we slightly refine the bounds of eqs. \eqref{eq:dv_bounds} and \eqref{eq:dchi_bounds}.

We start by looking at the velocity-dependence of $\Delta v^\mu_{(n,k,l)}$. We find the structure
\begin{subequations}
\begin{equation}
  \Delta v^\mu_{(n,k,l)}\sim 
  v^{i} (\hat{a}\cdot \hat{p}_\mathrm{rel})^j\quad\quad \forall\,\mu\in \{\hat V,\hat p_\mathrm{rel},\hat b,\hat l\},
\end{equation}
shared by every component $\mu\in \{\hat V,\hat p_\mathrm{rel},\hat b,\hat l\}$. Here $i$ and $j$ obey
\begin{align}\label{eq:conditions_structure_dv}
  &0\leq j\leq l
\\[5pt]\nonumber
  \text{and }&\begin{cases}
    \begin{aligned}
    &2(k+l)+1&&\lesssim \quad i\quad \leq&& 2(n+k+l)-1+\mathrm{ceil}_2(j), \quad &&(k+l) \text{  odd}
    \\
    &2(k+l)&&\lesssim \quad i\quad \leq&& 2(n+k+l)+\mathrm{floor}_2(j), \quad &&(k+l) \text{  even}
  \end{aligned}
  \end{cases}
\end{align}
\end{subequations}
$\mathrm{ceil}_2(j)$ and $\mathrm{floor}_2(j)$ denote the floor and ceiling functions to the nearest integer multiple of 2. 
Although other dot-products appear, only powers of $(\hat{a}\cdot \hat{p}_\mathrm{rel})$ affect the bounds on powers of velocity.

For a probe which is not Kerr \footnote{For a Kerr probe, described by the Wilson coefficients of eqs. \eqref{eq:wilsoncoeff_EBbasis} and \eqref{eq:wilsoncoeff_R2_Kerr}, the $(n\geq 2,k=0,l=4)$ results continue to follow eq. \eqref{eq:conditions_structure_dv}.}, the $(n\geq 2,k=0,l=4)$ results further deviate from  eq. \eqref{eq:conditions_structure_dv}. These pieces of $\Delta v^\mu$ carry an additional factor of $\gamma^2$, and take the form
\begin{equation}\label{eq:dv_exception}
  \Delta v^\mu_{(n,k,l)}\sim 
  \gamma^2v^{i} (\hat{a}\cdot \hat{p}_\mathrm{rel})^j, \quad\quad\quad (n\geq 2,k=0,l=4),
\end{equation}
with bounds on $i$ also differing from eq. \eqref{eq:conditions_structure_dv},
\begin{equation}
  2(k+l)\lesssim i\leq 2(n+k+l)+2+\mathrm{floor}_2(j).
\end{equation}
We note that these $\gamma$-factors do not appear for Kerr probes.

The structure of $\Delta \chi_{(n,k,l)}^\mu$ is a bit more complicated, yet similar to $\Delta v_{(n,k,l)}^\mu$. Notably, the $\hat V$ and $\{\hat b,\hat p_\mathrm{rel},\hat l\}$ components take different forms
\begin{subequations}
\begin{equation}
  \Delta \chi_{(n+k+l)}^\mu\sim\begin{cases}
    \gamma^{\mathrm{floor}_2(l)}v^{i_0}(\hat{a}\cdot \hat{p}_\mathrm{rel})^j, \quad\quad \mu=\hat V\\
    \gamma^{\mathrm{floor}_2(l)}v^{i_s}(\hat{a}\cdot \hat{p}_\mathrm{rel})^j,\quad\quad\mu=\{\hat b,\hat p_\mathrm{rel}, \hat l\}
  \end{cases},
\end{equation}
where $0\leq j\leq l$ and
\begin{align}\label{eq:conditions_structure_dchi}
  \text{$(k+l)$ even}\quad
  &\begin{cases}
    2(k+l)+2\quad\lesssim\quad i_0\quad\leq\quad 2 (n+k+l) + \mathrm{floor}_2 (j) + \mathrm{floor}_2 (l) \\
    2(k+l)+1\quad\lesssim \quad i_s\quad\leq\quad 2 (n+k+l) - 1 + \mathrm{ceil}_2 (j) + \mathrm{floor}_2 (l)
  \end{cases}\\\nonumber
  \text{$(k+l)$ odd}\quad
  &\begin{cases}
    2(k+l)+1\quad \lesssim\quad  i_0\quad \leq\quad  2 (n+k+l) - 1 + \mathrm{ceil}_2 (j) + \mathrm{floor}_2 (l)\\
    2(k+l)+2\quad\lesssim\quad  i_s\quad\leq\quad  2 (n+k+l) + \mathrm{floor}_2 (j) + \mathrm{floor}_2 (l)
  \end{cases}
  .
\end{align}
\end{subequations}

There are a few exceptions to this rule; The $\hat V$ component for $(n,k=\text{even},l=3)$ has $j$-independent upper bounds on $i_0$,
\begin{equation}
  \hat V\text{ component @ } (n,k=\text{even},l=3)\quad \begin{cases}
      i_0\leq 2 n+1 + \mathrm{floor}_2 (l)=2n+3, \quad  & j>0\\
      i_0\leq 2 n - 1 + \mathrm{floor}_2 (l)=2n+1,\quad&j=0
    \end{cases}.
\end{equation}

\bibliographystyle{JHEP}
\bibliography{resummed.bib}

\end{document}